\documentclass[12pt]{article}
\usepackage{amsmath,amssymb,bm,epsf,epsfig,graphicx,hyperref,physics, cleveref, subfig, slashed, stmaryrd, tensor, float,longtable,booktabs}
\usepackage[backend=bibtex,natbib,style=numeric-comp,sorting=none,maxbibnames=99,doi=false,isbn=false,url=false]{biblatex}
\input epsf.sty
\usepackage[]{todonotes}
\usepackage{tikz}
\usetikzlibrary{arrows.meta, decorations.markings}

\usepackage[weather, alpine]{ifsym}

\def\ie{\begin{equation}\begin{aligned}}
\def\fe{\end{aligned}\end{equation}}
\def\NS{\mathrm{NS}}
\def\R{\mathrm{R}}
\def\F{\mathsf{F}}

\renewcommand{\title}[1]{\vbox{\center\LARGE{#1}}\vspace{5mm}}
\renewcommand{\author}[1]{\vbox{\center#1}\vspace{5mm}}
\newcommand{\address}[1]{\vbox{\center\em#1}}
\newcommand{\email}[1]{\vbox{\center\tt#1}\vspace{5mm}}

\begin{document}

\unitlength = .8mm

\begin{titlepage}

\begin{center}

\hfill \\
\hfill \\
\vskip 1cm

\title{All-Genus 2D \texorpdfstring{$\mathcal{N}=1$}{N=1} Superconformal Blocks}

\author{Sam Christian, Yuchen Wang, Xi Yin, Yutai Zhang}

\address{Jefferson Physical Laboratory, Harvard University,
Cambridge, MA 02138 USA
}

\email{
    samchristian@fas.harvard.edu, yuchen\_wang@fas.harvard.edu, xiyin@fas.harvard.edu, yutaizhang@fas.harvard.edu}

\end{center}

\abstract{We develop a numerical algorithm for two dimensional $\mathcal{N}=1$ superconformal blocks at arbitrary genera, in arbitrary plumbing channels, and for arbitrary spin structures. The algorithm utilizes two commuting copies of Virasoro subalgebras of the direct sum of the $\mathcal{N}=1$ superconformal algebra with an auxiliary free fermion algebra, which allows us to reduce the $\mathcal{N}=1$ superconformal blocks to Virasoro conformal blocks.}

\vfill

\end{titlepage}

\eject

\begingroup
\hypersetup{linkcolor=black}

\endgroup

\section{Introduction}

Superconformal blocks are the central objects for computing correlators of superconformal field theories (SCFTs). Given the superconformal block on a spin Riemann surface, together with the conformal data of the SCFT, namely the conformal weights of the superconformal primaries and the structure constants, correlators of a two dimensional $\mathcal{N}=1$ SCFT on that spin Riemann surface can be computed through the superconformal block decomposition. 

Direct evaluation of superconformal blocks by summing over descendants is computationally expensive, and calculations beyond the first few levels require more efficient numerical algorithms. On spin Riemann surfaces of genus $g=0$ or $g=1$, algorithms for numerically computing the superconformal blocks are well developed. Analogously to the Zamolodchikov recursion relation for the sphere 4-point Virasoro conformal block \cite{Zamolodchikov:1984eqp,Zamolodchikov:1987avt}, the sphere 4-point superconformal block can be computed using recursion relations in the central charge $c$ or in the intermediate conformal weights $h$. These recursion relations have been worked out for both the case with all-NS internal states \cite{Hadasz:2006qb,Hadasz:2007nt,Belavin:2007zz} and that with Ramond internal states \cite{Hadasz:2008dt,Suchanek:2010kq}. Similar to the case of Virasoro conformal blocks \cite{Cho:2017oxl}, the $c$-recursion relation of superconformal blocks can be generalized to arbitrary plumbing channels and arbitrary genera as long as all intermediate states are in the NS sector \cite{Belavin:2018hfm}. Meanwhile for the blocks that involve Ramond intermediate states, the $h$-recursion relations can only be generalized to genus 1 \cite{Hadasz:2012im, Collier:2026bhh} and the $c$-recursion relation cannot be generalized beyond tree level, as the large-$h$ or large-$c$ limit of these Ramond blocks does not exist at higher genus, and hence the regular part of the superconformal block cannot be read off from such limits. Algorithms for numerically computing these higher-genus Ramond blocks are absent in the literature.

In this work, we present a numerical algorithm for these higher-genus Ramond blocks. The algorithm is based on the fact that the $\mathcal{N}=1$ superconformal algebra $\mathsf{SCA}$, direct summed with an auxiliary free fermion algebra $\mathsf{F}$, admits two copies of commuting Virasoro subalgebras \cite{Bonelli:2011jx, Bonelli:2011kv, Belavin:2011sw, Schomerus:2012se, Bershtein:2014yia,Bershtein:2016uov}. These subalgebras were first discovered in the context of the AGT relation \cite{Alday:2009aq} and have been used in the context of the super-Liouville/double Liouville correspondence \cite{Hadasz:2013dza}. Using this $\mathsf{Vir}\oplus \mathsf{Vir}$ subalgebra, we can reduce the superconformal blocks to  Virasoro conformal blocks, which can be computed numerically as a series in the plumbing parameters using the $c$-recursion relation developed in \cite{Cho:2017oxl}.

The rest of this paper is organized as follows. In Section \ref{sec:scblock}, we define the superconformal blocks using the plumbing fixture. In Section \ref{sec:2Vir}, we review the details of the $\mathsf{Vir}\oplus \mathsf{Vir}$ subalgebra and the branching rules of $\mathsf{SCA}\oplus \mathsf{F}$ supermultiplets. In Section \ref{sec:blockdecomp}, we derive the relation between superconformal blocks and Virasoro blocks for blocks without a certain type of obstruction which we call odd Ramond loops. Section \ref{sec:oddloop} extends the construction to blocks with odd Ramond loops using an auxiliary fermion insertion. We present numerical tests for the algorithm in Section \ref{sec:numerics} and discuss applications and future directions in Section \ref{sec:disc}. The code for our algorithm and numerical tests is available on GitHub\footnote{\url{https://github.com/Yuchen-Wang-Harvard/sc_blocks}}.

\section{The superconformal blocks}

\label{sec:scblock}

A punctured Riemann surface can be constructed by gluing 3-punctured Riemann spheres using the plumbing fixture, which takes two punctures with local coordinates $w_1, w_2$ and identifies a pair of annular regions around the punctures using
\ie 
w_1w_2=q,\quad |q|\leq 1,
\fe
where $q$ is a complex plumbing parameter. A genus-$g$, $n$-punctured Riemann surface can be constructed by plumbing $(3g-3+n)$ pairs of punctures, and its $(3g-3+n)$ complex moduli are given by those $(3g-3+n)$ plumbing parameters. 

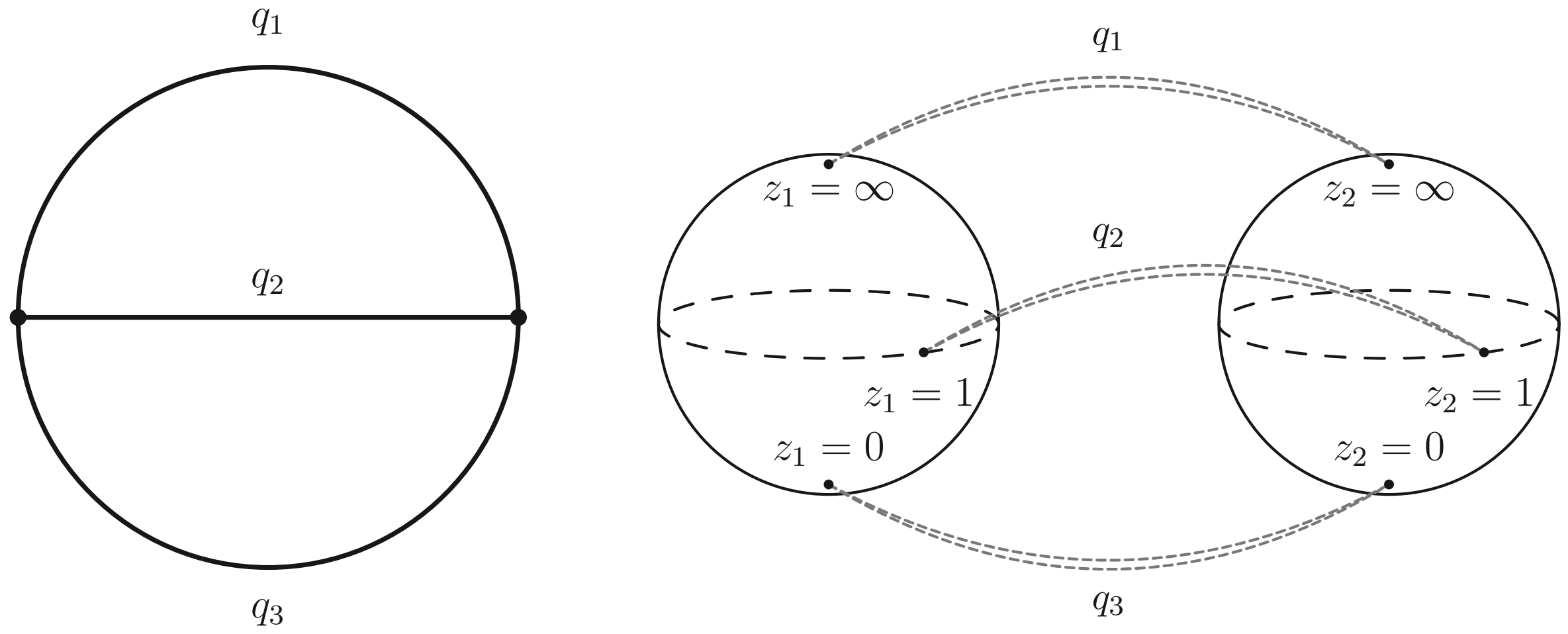
\begin{figure}
\centering
\begin{tikzpicture}[
  x=1cm, y=1cm,
  line cap=round, line join=round, color=black!91,
  sphere/.style={black!91, line width=0.8pt},
  equator/.style={sphere, dash pattern=on 5.6pt off 5.6pt},
  tube/.style={draw=black!53, line width=0.8pt, dash pattern=on 2.4pt off 1.6pt},
  every node/.style={inner sep=2pt, font=\fontsize{12}{14}\selectfont},
  graph label/.style={font=\fontsize{12}{14}\selectfont},
  tube label/.style={font=\small}
]

\draw[line width=1.35pt] (2.5,0) circle[radius=2.5];
\draw[line width=1.35pt] (0,0) -- (5,0);
\fill (0,0) circle[radius=2.4pt];
\fill (5,0) circle[radius=2.4pt];
\node[graph label] at (2.50,2.95) {$q_1$};
\node[graph label] at (2.50,0.35) {$q_2$};
\node[graph label] at (2.50,-2.95) {$q_3$};

\foreach \x in {8.1,13.7} {
  \draw[sphere] (\x,-0.07) circle[radius=1.7];
  \draw[equator] (\x,-0.07) ellipse[x radius=1.7,y radius=0.34];
}

\coordinate (Linf)  at (8.1, 1.53);
\coordinate (Rinf)  at (13.7,1.53);
\coordinate (Lone)  at (9.05,-0.35);
\coordinate (Rone)  at (14.65,-0.35);
\coordinate (Lzero) at (8.1,-1.67);
\coordinate (Rzero) at (13.7,-1.67);

\draw[tube] (Linf) .. controls (9.97,2.69) and (11.83,2.69) .. (Rinf);
\draw[tube] (Linf) .. controls (9.97,2.57) and (11.83,2.57) .. (Rinf);

\draw[tube] (Lone) .. controls (10.92,0.81) and (12.78,0.81) .. (Rone);
\draw[tube] (Lone) .. controls (10.92,0.69) and (12.78,0.69) .. (Rone);

\draw[tube] (Lzero) .. controls (9.97,-2.80) and (11.83,-2.80) .. (Rzero);
\draw[tube] (Lzero) .. controls (9.97,-2.68) and (11.83,-2.68) .. (Rzero);

\foreach \p in {Linf,Rinf,Lone,Rone,Lzero,Rzero}
  \fill (\p) circle[radius=1.4pt];

\node at (8.10,1.23) {$z_1=\infty$};
\node at (13.70,1.23) {$z_2=\infty$};
\node at (9.00,-0.78) {$z_1=1$};
\node at (14.60,-0.78) {$z_2=1$};
\node at (8.10,-1.32) {$z_1=0$};
\node at (13.70,-1.32) {$z_2=0$};

\node[tube label] at (10.90,2.77) {$q_1$};
\node[tube label] at (10.90,0.81) {$q_2$};
\node[tube label] at (10.90,-2.87) {$q_3$};

\end{tikzpicture}

    \caption{The genus two theta channel. The left picture shows the trivalent graph $\Gamma$, and the right picture shows the plumbing construction for a genus two Riemann surface. The tubes implement the plumbing identifications $\frac{1}{z_1z_2}=q_1$, $(z_1-1)(z_2-1)=q_2$, and $z_1z_2=q_3$.}
    \label{fig:thetachannel}
\end{figure}

On a 3-punctured Riemann sphere with global complex coordinate $z$, we take the three punctures to be located at $z=\infty, 1, 0$. Although the plumbing fixture allows arbitrary local coordinates $w_\infty, w_1, w_0$ at these punctures, throughout this paper we take their respective local coordinates to be
\ie 
w_\infty = \frac{1}{z},\quad w_1=z-1,\quad w_0=z,
\fe
and focus on Riemann surfaces that can be constructed this way.\footnote{Although it is generally an open problem if every punctured Riemann surface can be constructed by plumbing together 3-punctured spheres with these local coordinate maps, the genus two free energy computation we performed in \cite{Christian:new} suggests that it is true for genus two closed Riemann surfaces.} One can then describe a punctured Riemann surface using a trivalent graph $\Gamma$, together with a $q$-parameter for each internal edge. The vertices of $\Gamma$ correspond to the 3-punctured spheres, while the edges connect pairs of plumbed punctures. We call this graph $\Gamma$ a plumbing channel. An example of such a construction, the genus two theta channel, is shown in figure \ref{fig:thetachannel}.

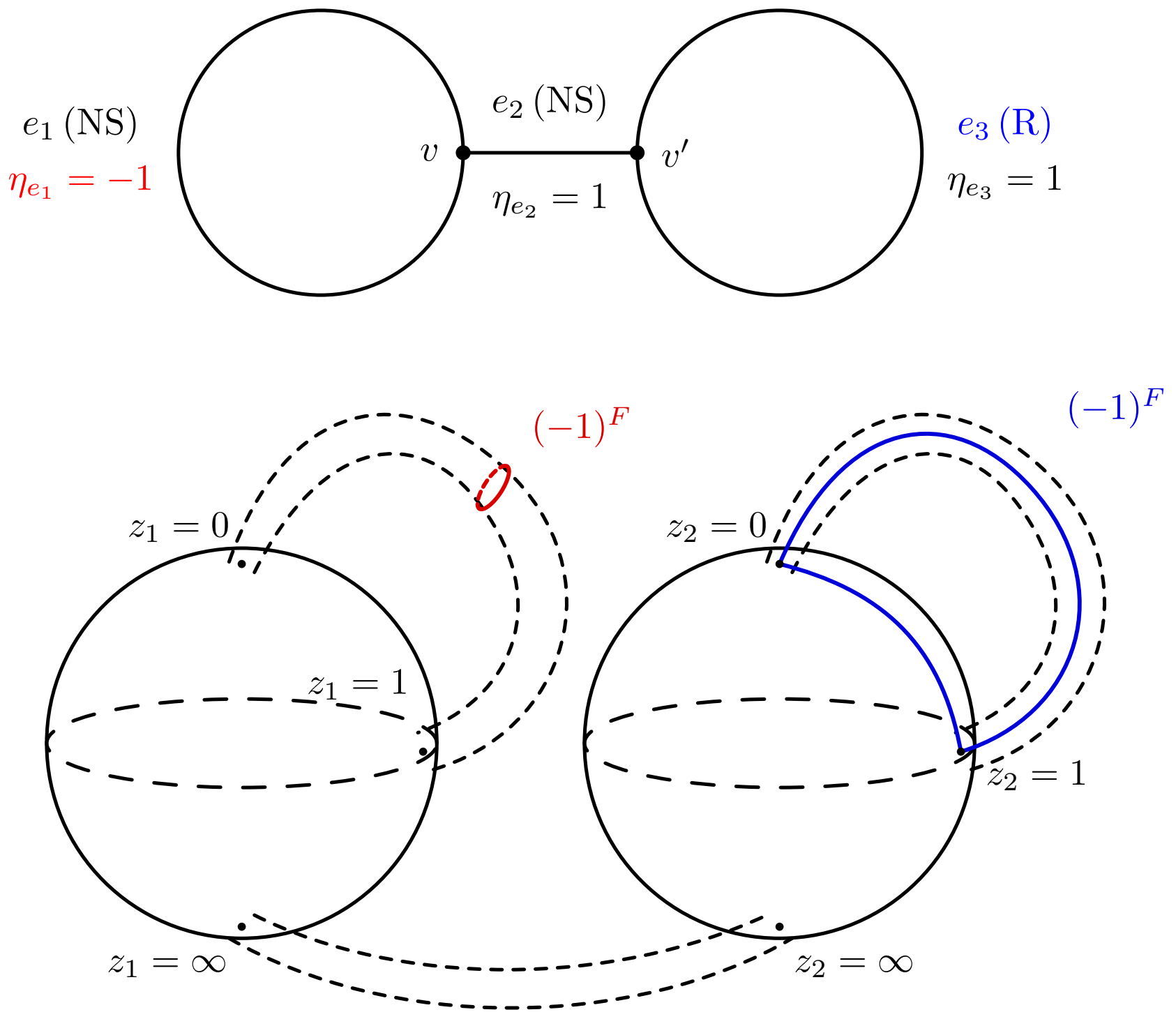
\begin{figure}
\begin{center}
\begin{tikzpicture}[
  x=1cm, y=1cm,
  line width=0.95pt, line cap=round, line join=round,
  tube/.style={dashed},
  every node/.style={inner sep=2pt, font=\fontsize{10}{12}\selectfont},
  twist1/.style={draw=red!85!black, line width=1.1pt},
  twist2/.style={draw=blue!85!black, line width=1.1pt},
  red label/.style={text=red!85!black},
  blue label/.style={text=blue!85!black}
]

\draw (2.85,8.15) circle[radius=1.35];
\draw (7.20,8.15) circle[radius=1.35];
\draw (4.20,8.15) -- (5.85,8.15);
\foreach \x in {4.20,5.85}
  \fill[black] (\x,8.15) circle[radius=2pt];
\node[left=4pt]  at (4.20,8.15) {$v$};
\node[right=4pt] at (5.85,8.15) {$v'$};
\node[anchor=east,align=center] at (1.35,8.15)
  {$e_1\, (\mathrm{NS})$\\[3pt]$\color{red}{\eta_{e_1}=-1}$};
\node[above=6pt] at (5.025,8.15) {$e_2\, (\mathrm{NS})$};
\node[below=6pt] at (5.025,8.15) {$\eta_{e_2}=1$};
\node[anchor=west,align=center] at (8.70,8.15)
  {$\color{blue}{e_3\, (\mathrm{R})}$\\[3pt]$\eta_{e_3}=1$};

\foreach \x in {2.10,7.20} {
  \begin{scope}[shift={(\x,2.55)}]
    \draw (0,0) circle[radius=1.85];
    \draw[dash pattern=on 6pt off 6pt]
      (0,0) ellipse[x radius=1.85,y radius=0.42];
    \draw[tube] (-0.12,1.72)
      .. controls (0.42,3.27) and (1.55,3.48) .. (2.50,2.60)
      .. controls (3.55,1.62) and (3.10,0.10) .. (1.78,-0.26);
    \draw[tube] (0.12,1.62)
      .. controls (0.76,2.91) and (1.69,3.05) .. (2.27,2.25)
      .. controls (2.95,1.34) and (2.58,0.42) .. (1.67,0.10);
  \end{scope}
}

\draw[tube] (1.98,0.70)
  .. controls (3.50,-0.18) and (5.80,-0.18) .. (7.32,0.70);
\draw[tube] (2.22,0.92)
  .. controls (3.60,0.23) and (5.60,0.23) .. (7.08,0.92);

\begin{scope}[shift={(4.485,4.975)},rotate=56.7]
  \draw[twist1] (-0.24,0)
    arc[start angle=180,end angle=360,x radius=0.24,y radius=0.085];
  \draw[twist1,dash pattern=on 2pt off 1.5pt] (0.24,0)
    arc[start angle=0,end angle=180,x radius=0.24,y radius=0.085];
\end{scope}
\node[red label,above right] at (4.76,5.27) {$(-1)^F$};

\begin{scope}[shift={(7.20,2.55)}]
  \draw[twist2] (0,1.70)
    .. controls (0.59,3.09) and (1.62,3.27) .. (2.385,2.425)
    .. controls (3.24,1.48) and (2.84,0.26) .. (1.72,-0.08)
    .. controls (1.53,0.96) and (0.95,1.46) .. (0,1.70);
\end{scope}
\node[blue label,above right] at (9.83,5.45) {$(-1)^F$};

\foreach \x/\i in {2.10/1,7.20/2} {
  \begin{scope}[shift={(\x,2.55)}]
    \foreach \p in {(0,1.70),(1.72,-0.08),(0,-1.74)}
      \fill \p circle[radius=1.1pt];
    \node[above left] at (-0.03,1.82) {$z_{\i}=0$};
  \end{scope}
}
\node[above left]  at (3.77,2.88) {$z_1=1$};
\node[below right] at (9.07,2.47) {$z_2=1$};
\node[below left]  at (2.05,0.62) {$z_1=\infty$};
\node[below right] at (7.25,0.62) {$z_2=\infty$};

\end{tikzpicture}
\end{center}

    \caption{Specifying a spin structure on the Riemann surface constructed in the genus two glasses channel. The upper picture shows the labeled graph, while the lower picture shows the corresponding plumbing construction. The colored circles in the lower picture are the insertions of topological line operators $(-1)^F$: The red circle depicts a $(-1)^F$ insertion around the left plumbing tube, which corresponds to a spin lift sign $\eta_{e_1}=-1$. The blue circle depicts to a $(-1)^F$ insertion through the right handle and the right sphere, which corresponds to that the intermediate states propagating along $e_3$ being in the Ramond sector.}
    \label{spingraph}
\end{figure}

To specify a holomorphic spin structure on the Riemann surface, we assign a sign $\eta_e$ for each internal edge $e$ in $\Gamma$, and additionally label each edge of $\Gamma$ by Neveu-Schwarz (NS) or Ramond (R). In the $\mathcal{N}=1$ SCFT language, the spin lift sign $\eta_e$ specifies whether there is a line operator insertion $(-1)^F$ around the tube associated with edge $e$, while the NS/R label specifies the fermionic boundary conditions of the states propagating along the tube. At each vertex, only edges of types (NS, NS, NS) or (NS, R, R) can be connected together. We denote the set of the NS/R labels by $L$, and the data of the holomorphic spin structure is specified by $(L, \{\eta_e\})$. An example of this labelled graph is shown in figure \ref{spingraph}. Similarly, an anti-holomorphic spin structure can be specified by a set $\tilde{L}$ of NS/R labels and a set of spin lift signs $\tilde{\eta}_e$.

In the following discussions, the sets of all NS internal edges, R internal edges, (NS, NS, NS) vertices and (NS, R, R) vertices in the labeled graph $(\Gamma, L)$ are denoted by $E_{\NS}, E_{\R}, V_{\NS}$ and $V_{\R}$, respectively. In order to keep track of the Grassmann signs, we fix an order of all the vertices in $\Gamma$, specifically such that the (NS, NS, NS) vertices precede all the (NS, R, R) vertices. We also keep track of the order of the two half-edges in each edge, as well as the order of the three half-edges that connect to each vertex, with the first half-edge corresponding to the operator inserted at $z=\infty$, the second corresponding to $z=1$, and the third corresponding to $z=0$. On an (NS, R, R) sphere, we always take the NS half-edge to be in the first slot. 

Given a plumbing construction of a Riemann surface, one can define a Weyl frame on that Riemann surface characterized by the flat metric $ds^2 = 2dz d\bar{z}$ on each 3-punctured sphere. The CFT correlators in this Weyl frame can be computed by summing over a complete basis of intermediate states on the plumbing tubes. In a $\mathcal{N}=1$ SCFT, due to the holomorphic and anti-holomorphic superconformal symmetry, these intermediate states organize into tensor product of holomorphic and anti-holomorphic supermultiplets. We denote a holomorphic NS superconformal primary by $\phi$. A holomorphic Ramond supermultiplet has 2 ground states $w^\pm$ that are related by $G_0$:
\ie
G_0 w^\pm = i \beta e^{\mp i\pi/4} w^\mp,
\fe
where $\beta$ is defined by $\beta^2 = \frac{c}{24}-h$. We take $w^+$ to be Grassmann even, and hence $w^-$ will be Grassmann odd. A basis of states in a holomorphic supermultiplet is given by 
\ie
&\mathbf{L}_{-A}\phi \equiv L_{-n_1}\cdots L_{-n_\ell} G_{-r_1}\cdots G_{-r_p} \phi,\\
&\mathbb{L}_{-A} w^\pm \equiv L_{-n_1}\cdots L_{-n_\ell} G_{-r_1}\cdots G_{-r_p} w^\pm,
\fe
in the NS and R sectors, respectively. Here, $n_i\in \mathbb{Z}, r_j\in \mathbb{Z}+\nu$, and $n_1 \geq ... \geq n_\ell>0, r_1>...>r_p>0$, where $\nu=0$ in the R sector and $\nu=\frac{1}{2}$ in the NS sector. We further define the total level of $\mathbf{L}_{-A}, \mathbb{L}_{-A}$ by $|A|\equiv \sum_{i=1}^\ell n_i + \sum_{j=1}^{p} r_j$, and their Grassmann parity by $(-1)^A \equiv (-1)^{p}$. 

An (NS, NS) superconformal primary $\mathcal{O}(z,\bar{z})$ in a $\mathcal{N}=1$ SCFT admit a formal decomposition
\ie \label{primdecomp}
\mathcal{O}(z, \bar{z}) = \phi(z) \otimes \tilde{\phi}(\bar{z}),
\fe
and similar decomposition works for primaries in holomorphic and/or anti-holomorphic Ramond sector by replacing the corresponding $\phi, \tilde{\phi}$ by $w^\alpha, \tilde{w}^{\tilde{\alpha}}$. A basis of states of the SCFT is then given by the tensor product of $\mathbf{L}_{-A}\phi, \mathbb{L}_{-A}w^\alpha$ and $\widetilde{\mathbf{L}}_{-\tilde{A}}\tilde{\phi}, \widetilde{\mathbb{L}}_{-\tilde{A}}\tilde{w}^\alpha$.

After replacing each pair of plumbed punctures by a pair of local operators, we can reduce the SCFT correlator to a product of 3-point functions on the spheres. For simplicity, we will focus on 0-point superconformal blocks in the rest of this paper, which enter the computation of SCFT partition functions. We will also focus on the case where the holomorphic and anti-holomorphic spin structures are diagonal, namely $L=\tilde{L}$ and $\eta_e = \tilde{\eta}_e$. Generalization to arbitrary blocks with external states and non-diagonal spin structures is straightforward. On a spin Riemann surface specified by the plumbing parameters $(\Gamma,\{q_e\})$ and diagonal spin structure $(L, \{\eta_e\})$, the partition function can be computed by summing over contributions of intermediate operators propagating on each plumbing tube:
\ie
Z_{\Gamma, L}&(q_e, \bar{q}_e, \eta_e) = \sum_{\text{primaries}} \sum_{\{A,\alpha\}} (-1)^{K_\Gamma(\epsilon+\tilde{\epsilon})} \prod_{e\in E_\NS \cup E_\R} q_e^{h_e+|A_{e_1}|}\bar{q}_e^{\tilde{h}_e+|\tilde{A}_{e_1}|}\eta_e^{\epsilon_{e_1}+\tilde{\epsilon}_{e_1}} \\ 
&\quad\times \prod_{e'\in E_\NS} (-1)^{\tilde{\epsilon}_{e_1'} \epsilon_{e_2'}} \mathbf{G}_{h_{e'}}^{A_{e_1'}A_{e_2'}} \mathbf{G}_{\tilde{h}_{e'}}^{\tilde{A}_{e_1'}\tilde{A}_{e_2'}} \prod_{e''\in E_\R} (-1)^{\tilde{\epsilon}_{e_1''} \epsilon_{e_2''}} \mathbb{G}_{\beta_{e''}}^{A_{e_1''}\alpha_{e_1''},A_{e_2''}\alpha_{e_2''}} \mathbb{G}_{\tilde{\beta}_{e''}}^{\tilde{A}_{e_1''}\tilde{\alpha}_{e_1''},\tilde{A}_{e_2''}\tilde{\alpha}_{e_2''}}\\
&\quad\times \prod_{v\in V_\NS}\langle \langle \mathbf{L}_{-A_{v_1}} \phi_{v_1} \otimes \widetilde{\mathbf{L}}_{-\tilde{A}_{v_1}}\tilde{\phi}_{v_1}| \mathbf{L}_{-A_{v_2}} \phi_{v_2} \otimes \widetilde{\mathbf{L}}_{-\tilde{A}_{v_2}}\tilde{\phi}_{v_2}(1,1)|\mathbf{L}_{-A_{v_3}} \phi_{v_3} \otimes \widetilde{\mathbf{L}}_{-\tilde{A}_{v_3}}\tilde{\phi}_{v_3}\rangle\\
&\quad\times \prod_{v\in V_\R}\langle \langle \mathbf{L}_{-A_{v_1}} \phi_{v_1} \otimes \widetilde{\mathbf{L}}_{-\tilde{A}_{v_1}}\tilde{\phi}_{v_1}| \mathbb{L}_{-A_{v_2}} w_{v_2}^{\alpha_{v_2}} \otimes \widetilde{\mathbb{L}}_{-\tilde{A}_{v_2}}\tilde{w}_{v_2}^{\tilde{\alpha}_{v_2}}(1,1)|\mathbb{L}_{-A_{v_3}} w_{v_3}^{\alpha_{v_3}} \otimes \widetilde{\mathbb{L}}_{-\tilde{A}_{v_3}}\tilde{w}_{v_3}^{\tilde{\alpha}_{v_3}}\rangle.
\fe

Let us unpack the notations used in the formula here. The first sum is over all superconformal primaries in the SCFT that propagate on each edge, and $\phi,\tilde{\phi}, w, \tilde{w}$ are the holomorphic and anti-holomorphic primaries appearing in their decomposition \eqref{primdecomp}. The second sum is over all displayed descendant labels $A$ on each half-edge and all ground state labels $\alpha$ on each Ramond half-edge, subject to the conditions
\ie \label{selection-sewing}
&|A_{e_1}| = |A_{e_2}|,\quad |\tilde{A}_{e_1}|=|\tilde{A}_{e_2}|,\quad e\in E_{\NS}\cup E_\R,\\
&(-1)^{A_{e_1}} = (-1)^{A_{e_2}} ,\quad e\in E_\NS,\\
&(-1)^{A_{e_1}+\alpha_{e_1}} = (-1)^{A_{e_2}+\alpha_{e_2}},\quad e\in E_\R.
\fe
The two half-edges in edge $e$ are labeled $e_1, e_2$, and the three half-edges connected to a vertex $v$ are labeled $v_1, v_2, v_3$, both according to their aforementioned order. 

We also define $\epsilon_{v_i}$ to be the total Grassmann parity of the holomorphic part of the corresponding descendant, namely $\mathbf{L}_{-A_{v_i}}\phi_{v_i}$ in the NS sector or $\mathbb{L}_{-A_{v_i}}w_{v_i}^{\alpha_{v_i}}$ in the R sector. Similarly, $\tilde{\epsilon}_{v_i}$ is defined to be the total Grassmann parity of the antiholomorphic part. The sign $(-1)^{K_\Gamma(\epsilon+\tilde{\epsilon})}$ is the Grassmann sign that arises from moving the operator order from the pairwise insertion order to the displayed order in the 3-point functions. 

More explicitly, denoting the full operator insertion by $\Phi$, namely $\Phi=\mathbf{L}_{-A}\phi\otimes \widetilde{\mathbf{L}}_{-\tilde{A}}\tilde{\phi}$ for the NS sector and $\Phi=\mathbb{L}_{-A}w^\alpha\otimes \widetilde{\mathbb{L}}_{-\tilde{A}}\tilde{w}^{\tilde{\alpha}}$ for the R sector, $(-1)^{K_\Gamma(\epsilon+\tilde{\epsilon})}$ is the Grassmann sign that arises from moving the operators from the pairwise insertion order $\prod_{e\in E_\NS\cup E_\R}\Phi_{e_1}\Phi_{e_2}$ to the order $\prod_{v\in V_\NS}\Phi_{v_1}\Phi_{v_2}\Phi_{v_3}\prod_{v\in V_\R}\Phi_{v_1}\Phi_{v_2}\Phi_{v_3}$ that appeared in the 3-point functions. In this language, this Grassmann sign is given by
\ie \label{Ksign}
K_\Gamma(\epsilon+\tilde{\epsilon})=\sum_{(v_i, w_j)}(\epsilon_{v_i}+\tilde{\epsilon}_{v_i})(\epsilon_{w_j}+\tilde{\epsilon}_{w_j})\quad (\text{mod }2),
\fe 
where the sum is over pairs $(v_i, w_j)$ of half-edges whose order is swapped between the pairwise insertion order and the 3-point function order. 

Lastly, the matrices $(\mathbf{G}_h)^{AB}, (\mathbb{G}_\beta)^{A\alpha, B\alpha'}$ are the inverse Gram matrices in the NS and R sectors, defined to be the inverses of
\ie
(\mathbf{G}_h)_{AB}=\langle\langle \mathbf{L}_{-A}\phi_h|\mathbf{L}_{-B}\phi_h\rangle,\quad (\mathbb{G}_\beta)_{A\alpha, B\alpha'}=\langle\langle \mathbb{L}_{-A}w_\beta^\alpha|\mathbb{L}_{-B}w_\beta^{\alpha'}\rangle.
\fe
Here, the subscripts $h, \beta$ label the conformal weights of the primaries, whose normalization is taken to be \ie
\langle\langle \phi_h|\phi_h\rangle =1,\quad \langle\langle w_\beta^+|w_\beta^{+} \rangle=1.\fe

This partition function can be decomposed in terms of holomorphic and anti-holomorphic conformal blocks. One key fact behind this decomposition is the factorization of 3-point functions. On an (NS, NS, NS) sphere, all 3-point functions of descendants can be reduced to either $\langle \phi_1 \phi_2 \phi_3\rangle $ or $\langle \phi_1 (G_{-\frac{1}{2}}\phi_2) \phi_3\rangle $ by superconformal Ward identities. We define the holomorphic normalized 3-point functions $\rho_a$ by
\ie \label{rhoNS}
\rho_0(\phi_1, \phi_2, \phi_3)=1,\quad \rho_1(\phi_1, G_{-\frac{1}{2}}\phi_2, \phi_3)=1,
\fe
where the operators $\phi_1, \phi_2, \phi_3$ are inserted at $\infty, 1, 0$, respectively, and the operator at the first slot is inserted using the local coordinate map $w=1/z$. The holomorphic normalized three-point functions $\rho_a(\mathbf{L}_{-A}\phi_1, \mathbf{L}_{-B}\phi_2,\mathbf{L}_{-C}\phi_3)$ of descendants are defined as linear functions in their three arguments by reducing them to $\rho_0(\phi_1, \phi_2, \phi_3)$, $\rho_1(\phi_1, G_{-\frac{1}{2}}\phi_2, \phi_3)$ using superconformal Ward identities, which are given explicitly in appendix \ref{app:conventions}. On an (NS, R, R) sphere, there are 4 independent 3-point functions, and we define 
\ie \label{rhoR}
&\rho^{(\sigma)}_0(\phi_1, w^+_2, w_3^+) = 1,\quad \rho^{(\sigma)}_0(\phi_1, w^-_2, w_3^-) = \sigma,\\
&\rho^{(\sigma)}_1(\phi_1, w^+_2, w_3^-) = 1,\quad \rho^{(\sigma)}_1(\phi_1, w^-_2, w_3^+) = i\sigma,
\fe
for $\sigma = \pm$. Again, the operators $\phi_1, w_2^\pm, w_3^\pm$ are inserted at $\infty, 1, 0$, respectively, and the normalized 3-point functions of descendants are defined through superconformal Ward identities. Analogously, one can define the anti-holomorphic normalized 3-point functions $\tilde{\rho}_a$ and $\tilde{\rho}_f^{(\sigma)}$ by the complex conjugated normalization and complex conjugated superconformal Ward identities. Namely one normalizes $\tilde{\rho}_1^{(\sigma)}(\tilde{\phi}, \tilde{w}^-, \tilde{w}^+)=-i\sigma$, and their superconformal Ward identities have opposite $i$ factors compared to the ones satisfied by the holomorphic $\rho$.

On an (NS, NS, NS) sphere, the 3-point function can be factorized into
\ie 
\langle \langle &\mathbf{L}_{-A_{1}} \phi_{1} \otimes \widetilde{\mathbf{L}}_{-\tilde{A}_{1}}\tilde{\phi}_{1}| \mathbf{L}_{-A_{2}} \phi_{2} \otimes \widetilde{\mathbf{L}}_{-\tilde{A}_{2}}\tilde{\phi}_{2}(1,1)|\mathbf{L}_{-A_{3}} \phi_{3} \otimes \widetilde{\mathbf{L}}_{-\tilde{A}_{3}}\tilde{\phi}_{3}\rangle\\
&\qquad = (-1)^{\tilde{\epsilon}_1\epsilon_2+\tilde{\epsilon}_1\epsilon_3+\tilde{\epsilon}_2\epsilon_3}C_{123}^{(a)}\\
&\qquad \qquad \times \rho_a(\mathbf{L}_{-A_{1}} \phi_{1}, \mathbf{L}_{-A_{2}} \phi_{2}, \mathbf{L}_{-A_{3}} \phi_{3})\tilde{\rho}_{\tilde{a}}(\widetilde{\mathbf{L}}_{-\tilde{A}_{1}}\tilde{\phi}_{1}, \widetilde{\mathbf{L}}_{-\tilde{A}_{2}}\tilde{\phi}_{2}, \widetilde{\mathbf{L}}_{-\tilde{A}_{3}}\tilde{\phi}_{3}),
\fe 
where $C_{123}^{(a)}$ is the structure constant of the SCFT. The labels $a, \tilde{a}$ are determined by
\ie
(-1)^{A_1+A_2+A_3}=(-1)^a,\quad 
(-1)^{a+\epsilon(\phi_1)+\epsilon(\phi_2)+\epsilon(\phi_3)}=(-1)^{\tilde{a}+\epsilon(\tilde{\phi}_1)+\epsilon(\tilde{\phi}_2)+\epsilon(\tilde{\phi}_3)},
\fe
where $\epsilon(\phi_i)$ is the Grassmann parity of the NS-sector holomorphic primary $\phi_i$. Namely, $a$ is defined to be the total Grassmann parity of the holomorphic lowering operators $\mathbf{L}_{-A_1}, \mathbf{L}_{-A_2}, \mathbf{L}_{-A_3}$, and $\tilde{a}$ is fixed by the condition that the total Grassmann parity of $\Phi_1, \Phi_2, \Phi_3$ add up to 0. Note that this equation also determines our convention for the structure constants. On an (NS, R, R) sphere, a similar factorization identity holds:
\ie
&\langle \langle \mathbf{L}_{-A_{1}} \phi_{1} \otimes \widetilde{\mathbf{L}}_{-\tilde{A}_{1}}\tilde{\phi}_{1}| \mathbb{L}_{-A_{2}} w_{2}^{\alpha_2} \otimes \widetilde{\mathbb{L}}_{-\tilde{A}_{2}}\tilde{w}_{2}^{\tilde{\alpha}_2}(1,1)|\mathbb{L}_{-A_{3}} w_{3}^{\alpha_3} \otimes \widetilde{\mathbb{L}}_{-\tilde{A}_{3}}\tilde{w}_{3}^{\tilde{\alpha}_3}\rangle\\
& \qquad = (-1)^{\tilde{\epsilon}_1\epsilon_2+\tilde{\epsilon}_1\epsilon_3+\tilde{\epsilon}_2\epsilon_3} \sum_{\sigma = \pm} C_{123}^{(f,\sigma)}\\
&\qquad \qquad\times \rho_f^{(\sigma)}(\mathbf{L}_{-A_{1}} \phi_{1}, \mathbb{L}_{-A_{2}} w_{2}^{\alpha_2}, \mathbb{L}_{-A_{3}} w_{3}^{\alpha_3})\tilde{\rho}_{\tilde{f}}^{(\sigma)}(\widetilde{\mathbf{L}}_{-\tilde{A}_{1}}\tilde{\phi}_{1}, \widetilde{\mathbb{L}}_{-\tilde{A}_{2}}\tilde{w}_{2}^{\tilde{\alpha}_2}, \widetilde{\mathbb{L}}_{-\tilde{A}_{3}}\tilde{w}_{3}^{\tilde{\alpha}_3}),
\fe
where $C_{123}^{(f,\sigma)}$ is the structure constant, and $f, \tilde{f}$ are determined by
\ie 
(-1)^{A_1+A_2+A_3+\alpha_2+\alpha_3}=(-1)^f,\quad (-1)^{f+\epsilon(\phi_1)}=(-1)^{\tilde{f}+\epsilon(\tilde{\phi}_1)}.
\fe

Using this factorization, the partition function can be decomposed as
\ie
Z_{\Gamma, L}(q_e, \bar{q}_e, \eta_e) = \sum_{\text{primaries}} \sum_{\{a, f, \sigma\}}&(-1)^{S_\Gamma(\epsilon,\tilde{\epsilon})} \prod_{v\in V_\NS} C_{v_1v_2v_3}^{(a_v)} \prod_{v'\in V_\R} C_{v_1'v_2'v_3'}^{(f_{v'},\sigma_{v'})} \prod_{e\in E_\NS\cup E_\R} q_e^{h_e} \bar{q}_e^{\tilde{h}_e}\\
&\quad \times \mathbf{F}^{(f, \sigma, a)}_{\Gamma, L}(h_e, \beta_e; q_e, \eta_e) \widetilde{\mathbf{F}}^{(\tilde{f}, \sigma, \tilde{a})}_{\Gamma, L}(\tilde{h}_e, \tilde{\beta}_e; \bar{q}_e, \eta_e),
\fe
where the second sum goes over all labels $a_v$ on each (NS, NS, NS) vertex and all labels $\sigma_v$ on each (NS, R, R) vertex. $h_e, \tilde{h}_e$ are the holomorphic and anti-holomorphic conformal weights of the primary that propagates on edge $e$. The sign $S_\Gamma(\epsilon, \tilde{\epsilon})$ is defined by
\ie \label{sgamma}
p_v=\begin{cases}
a_v + \sum_{i=1}^3 \epsilon(\phi_{v_i}),\quad v\in V_\NS,\\
f_v + \epsilon(\phi_{v_1}),\quad v\in V_\R,
\end{cases}\quad S_\Gamma(\epsilon, \tilde{\epsilon}) =\sum_{v<v'} p_vp_{v'}\quad (\text{mod }2),
\fe
and the sum is over pairs of vertices $(v,v')$ such that $v$ precedes $v'$ in the aforementioned ordering of all vertices. Finally, $\mathbf{F}, \tilde{\mathbf{F}}$ are the holomorphic and anti-holomorphic superconformal blocks with internal weights $\{h_e\}_{e\in E_{\NS}}, \{\beta_e\}_{e\in E_\R}$. The holomorphic block can be computed by
\ie \label{scblock}
\mathbf{F}^{(f, \sigma, a)}_{\Gamma, L}(h_e, \beta_e; q_e, \eta_e) &= \sum_{\{A, \alpha\}}(-1)^{K_{\Gamma}(\epsilon)} \prod_{e\in E_\NS\cup E_\R} q_e^{|A_{e_1}|}\eta_e^{\epsilon_e}\\
&\times \prod_{e'\in E_\NS} \mathbf{G}_{h_{e'}}^{A_{e_1'}A_{e_2'}} \prod_{e''\in E_\R} \mathbb{G}_{\beta_{e''}}^{A_{e_1''}\alpha_{e_1''},A_{e_2''}\alpha_{e_2''}}\\
&\times \prod_{v \in V_\NS} \rho_{a_v}(\mathbf L_{-A_{v_1}}\phi_{v_1},\mathbf L_{-A_{v_2}}\phi_{v_2},\mathbf L_{-A_{v_3}}\phi_{v_3})\\
&\times\prod_{v'\in V_\R}\rho_{f_{v'}}^{(\sigma_{v'})}(\mathbf L_{-A_{v_1'}}\phi_{v_1'},\mathbb L_{-A_{v_2'}}w_{v_2'}^{\alpha_{v_2'}},\mathbb L_{-A_{v_3'}}w_{v_3'}^{\alpha_{v_3'}}),
\fe
where the sign $K_\Gamma(\epsilon)$ is defined by replacing each $\epsilon +\tilde{\epsilon}$ by $\epsilon$ in \eqref{Ksign}. The sum over descendant labels and Ramond ground state labels is subject to the conditions
\ie \label{selection-block}
&|A_{e_1}| = |A_{e_2}|,\quad e\in E_{\NS}\cup E_\R\\
&(-1)^{A_{v_1}+A_{v_2}+A_{v_3}}=(-1)^{a_v},\quad v\in V_\NS,\\
&(-1)^{A_{v_1}+A_{v_2}+A_{v_3}+\alpha_{v_2}+\alpha_{v_3}}=(-1)^{f_v},\quad v\in V_\R.
\fe
The anti-holomorphic conformal block $\tilde{\mathbf{F}}$ is computed analogously using the anti-holomorphic normalized 3-point functions $\tilde{\rho}_a, \tilde{\rho}_f^{(\sigma)}$. In the rest of this paper, we will develop an algorithm for this quantity $\mathbf{F}^{(f, \sigma, a)}_{\Gamma, L}(h_e, \beta_e; q_e, \eta_e)$.

\section{The double Virasoro subalgebra}

\label{sec:2Vir}

Let us denote the $\mathcal{N}=1$ superconformal algebra with level shift $\nu\in \left\{0,\frac{1}{2}\right\}$ by $\mathsf{SCA}_\nu$, with $\nu=0$ corresponding to the Ramond algebra and $\nu=\frac{1}{2}$ corresponding to the Neveu-Schwarz algebra. The key fact we utilize is that, if we direct sum $\mathsf{SCA}_\nu$ with an auxiliary free fermion algebra $\mathsf{F}_\nu$ with the same shift $\nu$:
\ie
\{\psi_r, \psi_s\} &= \delta_{r+s,0},\quad \{\psi_r, G_s\}=[\psi_r, L_n] = 0,
\fe
the enlarged algebra $\mathsf{SCA}_\nu \oplus \mathsf{F}_{\nu}$ admits two commuting Virasoro subalgebras \cite{Belavin:2011sw, Schomerus:2012se, Bershtein:2014yia,Bershtein:2016uov}. 

The generators of the two Virasoro subalgebras can be written down explicitly as 
\ie
& L_n^{(1)}=\frac{b^{-1} L_n-\left(b^{-1}+2 b\right) L_n^{\mathrm{F}}+U_n}{b^{-1}-b}, \quad L_n^{(2)}=\frac{b L_n-\left(b+2 b^{-1}\right) L_n^{\mathrm{F}}+U_n}{b-b^{-1}},
\fe
where $b$ is defined by the $\mathsf{SCA}$ central charge $c= \frac{3}{2}+3\left(b+b^{-1}\right)^2$, and we have assumed that $b\neq 0,\pm 1$. The operators $L_n^{\mathrm{F}}, U_n$ are defined by
\ie\label{eq:fermionVirasoro}
L_n^{\mathrm{F}} &= \frac{1}{2} \sum_{r\in \mathbb{Z}+\nu} r:\psi_{n-r}\psi_r: + \frac{1-2\nu}{16}\delta_{n,0},\quad U_n = \sum_{r\in \mathbb{Z}+\nu} \psi_{n-r} G_r.
\fe
These generators $L_n^{(1,2)}$ satisfy $L_n^{(1)}+L_n^{(2)}=L_n+L_n^{\mathrm{F}}$, and the two central charges of the two Virasoro algebras are given by
\ie \label{c12}
c^{(1)}=1+\frac{3 Q^2}{1-b^2}, \quad c^{(2)}=1+\frac{3 Q^2}{1-b^{-2}},
\fe
where the Liouville background charge $Q$ is defined such that $Q=b+b^{-1}$.

Let $\mathcal{M}_{\NS}(h), \mathcal{M}_{\R}(\beta)$ denote the tensor products of the auxiliary fermion Fock space with the corresponding SCFT Verma module. More explicitly, let us denote the two R-sector ground states for the free fermion by $u^\mathsf{a},\ \mathsf{a}=0,1$, which satisfy
\ie
\psi_0 u^{\mathsf{a}} = \frac{1}{\sqrt{2}} u^{1-\mathsf{a}}.
\fe
The highest weight states in $\mathcal{M}_{\NS}(h), \mathcal{M}_{\R}(\beta)$ are then respectively given by $\mathbf{1}\otimes \phi$ and $u^{\mathsf{a}}\otimes w^\alpha$, and the full modules $\mathcal{M}_{\NS}(h), \mathcal{M}_{\R}(\beta)$ are given by acting with $L_{-n}, G_{-r},\psi_{-r}$ with $n,r>0$ on these highest weight states. 

Given the conformal weight $h$ of an NS primary or $\beta$ of a Ramond ground state, we define its Liouville momentum $P$ by using $P^2=\frac{1}{4}Q^2-2h$ in the NS sector and $P^2=2\beta^2$ in the R sector, where the branch choice for $P$ is discussed in more detail in Appendix \ref{app:freefield}. For Liouville momentum $P\notin \frac{1}{2}(b\mathbb{Z}+b^{-1}\mathbb{Z})$, these representations have branching rules \cite{Bershtein:2014yia,Bershtein:2016uov}
\ie \label{branchingrules}
\mathcal{M}_{\NS}(h)&\cong\bigoplus_{n\in\frac12\mathbb Z} V_{c^{(1)}}(h_n^{(1)})\otimes V_{c^{(2)}}(h_n^{(2)}),\\
\mathcal{M}_{\R}(\beta)&\cong\bigoplus_{\alpha=0,1}\ \bigoplus_{n\in\frac12\mathbb Z+\frac14} V_{c^{(1)}}(h_n^{(1)})\otimes V_{c^{(2)}}(h_n^{(2)}),
\fe
Here, $V_c(h)$ is the Virasoro Verma module with highest weight $h$ and central charge $c$. In the rest of this paper, we denote the highest-weight vector in $V_{c^{(1)}}(h_n^{(1)})\otimes V_{c^{(2)}}(h_n^{(2)})$ by $v_n$ and $v_n^\alpha$ in NS and R sectors, respectively. Their conformal weights are
\ie \label{h12}
h_n^{(1)}=\frac{Q^2/4-(P+2nb)^2}{2(1-b^2)},\quad h_n^{(2)}=\frac{Q^2/4-(P+2nb^{-1})^2}{2(1-b^{-2})}.
\fe
The NS $\mathsf{Vir}\oplus \mathsf{Vir}$ primary $v_n$ has Grassmann parity $2n\bmod2$ and level $2n^2$ relative to $\mathbf{1}\otimes \phi$, and the R primary $v^\alpha_n$ has Grassmann parity $\alpha=0,1$ and level $2n^2-\frac18$ relative to $u^{0}\otimes w^+$. These $\mathsf{Vir\oplus Vir}$ primaries can be realized by a free field representation, which we review in appendix \ref{app:freefield}.

\section{Reduction of superconformal blocks}

\label{sec:blockdecomp}

The strategy of our algorithm is to first consider an enlarged block with the full supermultiplet $\mathcal{M}_\NS(h), \mathcal{M}_\R(\beta)$ of the enlarged algebra $\mathsf{SCA}\oplus \mathsf{F}$ propagating on each tube. This enlarged block $\widehat{\mathbf{F}}^{(f,\sigma,a)}_{\Gamma, L}$ can be computed in two different ways. One can take the basis $\Psi_{-\mathsf{A}}\mathbf{1}\otimes \mathbf{L}_{-A}\phi$ for states in $\mathcal{M}_\NS(h)$ and $\Psi_{-\mathsf{A}}u^{\mathsf{a}}\otimes \mathbb{L}_{-A}w^\alpha$ for $\mathcal{M}_{\R}(\beta)$, which allows one to write the enlarged block in terms of the original superconformal block and an auxiliary fermion block. Alternatively, one can compute the enlarged block using the decomposition \eqref{branchingrules}, which allows one to write the enlarged block as a sum of Virasoro conformal blocks. By comparing the result from the two approaches, the superconformal blocks can be related to the auxiliary fermion block and the Virasoro conformal blocks.

More precisely, we define the enlarged $\mathsf{SCA}\oplus \mathsf{F}$ block $\widehat{\mathbf{F}}^{(f,\sigma,a)}_{\Gamma, L}$ to be\footnote{The additional factor $(-1)^{\epsilon_e \epsilon_e'}$ follows from our convention \eqref{BPZtensor} for the BPZ inner product.}
\ie 
\widehat{\mathbf{F}}^{(f,\sigma,a)}_{\Gamma, L} &= \sum_{\{A,\alpha,\mathsf{A},\mathsf{a}\}} (-1)^{K_{\Gamma}(\epsilon+\epsilon')} \prod_{e\in E_\NS\cup E_\R} q_e^{|A_{e_1}|+|\mathsf{A}_{e_1}|}\eta_e^{\epsilon_e+\epsilon_e'}(-1)^{\epsilon_e\epsilon_e'}\\
&\times \prod_{e\in E_\NS} \mathbf{G}_{\F}^{\mathsf{A}_{e_1}\mathsf{A}_{e_2}}  \mathbf{G}_{h_e}^{A_{e_1}A_{e_2}} \prod_{e\in E_\R} \mathbb{G}_{\F}^{\mathsf{A}_{e_1}\mathsf{a}_{e_1},\mathsf{A}_{e_2}\mathsf{a}_{e_2}} \mathbb{G}_{\beta_e}^{A_{e_1}\alpha_{e_1},A_{e_2}\alpha_{e_2}}\\
&\times \prod_{v \in V_\NS} \hat{\rho}_{a_v}(\Psi_{-\mathsf{A}_{v_1}}\mathbf{1}\otimes \mathbf L_{-A_{v_1}}\phi_{v_1},\Psi_{-\mathsf{A}_{v_2}}\mathbf{1}\otimes \mathbf L_{-A_{v_2}}\phi_{v_2},\Psi_{-\mathsf{A}_{v_3}}\mathbf{1}\otimes \mathbf L_{-A_{v_3}}\phi_{v_3})\\
&\times\prod_{v\in V_\R}\hat{\rho}_{f_v}^{(\sigma_v)}(\Psi_{-\mathsf{A}_{v_1}}\mathbf{1}\otimes \mathbf L_{-A_{v_1}}\phi_{v_1},\Psi_{-\mathsf{A}_{v_2}} u^{\mathsf{a_{v_2}}}\otimes\mathbb L_{-A_{v_2}}w_{v_2}^{\alpha_{v_2}},\Psi_{-\mathsf{A}_{v_3}} u^{\mathsf{a_{v_3}}}\otimes\mathbb L_{-A_{v_3}}w_{v_3}^{\alpha_{v_3}})
\fe
where $\Psi_{-\mathsf{A}}$ denotes a chain of fermion creation operators $\psi_{-r_1}...\psi_{-r_k}$ with $r_1>...>r_k>0$, and $|\mathsf{A}|$ denotes its total level. $\epsilon$ and $\epsilon'$ denote the Grassmann parity of the $\mathsf{SCA}$ and the fermion part of the corresponding descendant, respectively. Just like in the $\mathsf{SCA}$ case, we define $(-1)^{\mathsf{A}}\equiv (-1)^k$. The sum is over all displayed descendant and ground state labels, subject to the conditions
\ie \label{Fselection}
&\mathsf{A}_{e_1} = \mathsf{A}_{e_2},\quad e\in E_{\NS}\cup E_\R\\
&\mathsf{a}_{e_1} = \mathsf{a}_{e_2},\quad e\in E_\R\\
&(-1)^{\mathsf{A}_{v_1}+\mathsf{A}_{v_2}+\mathsf{A}_{v_3}} = 1,\quad v\in V_{\NS}\\
&(-1)^{\mathsf{A}_{v_1}+\mathsf{A}_{v_2}+\mathsf{A}_{v_3}+\mathsf{a}_{v_2}+\mathsf{a}_{v_3}} = 1,\quad v\in V_{\R},
\fe
in addition to \eqref{selection-block}. The fermion Gram matrices are given by
\ie
(\mathbf{G}_{\F})_{\mathsf{A}\mathsf{B}} &= \langle \langle \Psi_{-\mathsf{A}}\mathbf{1}|\Psi_{-\mathsf{B}}\mathbf{1}\rangle = (-i)^{k\text{ mod }2} \delta_{\mathsf{AB}},\\
(\mathbb{G}_{\F})_{\mathsf{Aa},\mathsf{Bb}}&= \langle \langle \Psi_{-\mathsf{A}}u^{\mathsf{a}}|\Psi_{-\mathsf{B}}u^{\mathsf{b}}\rangle=(-i)^{k+\mathsf{a}\text{ mod }2}\delta_{\mathsf{AB}}\delta_{\mathsf{ab}},
\fe
where $k$ is the number of fermionic creation operators in $\Psi_{-\mathsf{A}}$, and the factors of $i$ come from our convention \eqref{BPZgenerator} of BPZ conjugation. The enlarged 3-point functions $\hat{\rho}_a, \hat{\rho}_f^{(\sigma)}$ are defined as
\ie 
 &\hat\rho_a(\Psi_{-\mathsf A}\mathbf1\otimes\mathbf L_{-A}\phi_1,
 \Psi_{-\mathsf B}\mathbf1\otimes\mathbf L_{-B}\phi_2,
 \Psi_{-\mathsf C}\mathbf1\otimes\mathbf L_{-C}\phi_3)\\
 &\qquad\qquad=(-1)^{\epsilon_1\epsilon_1'+\epsilon_2 \epsilon_3'}
 \rho_{\mathsf{F}}(\Psi_{-\mathsf A}\mathbf1,\Psi_{-\mathsf B}\mathbf1,
                                      \Psi_{-\mathsf C}\mathbf1)
 \rho_a(\mathbf L_{-A}\phi_1,\mathbf L_{-B}\phi_2,\mathbf L_{-C}\phi_3),\\
 &\hat\rho_f^{(\sigma)}(\Psi_{-\mathsf A}\mathbf1\otimes\mathbf L_{-A}\phi_1,
 \Psi_{-\mathsf B}u^{\mathsf b}\otimes\mathbb L_{-B}w_2^\alpha,
 \Psi_{-\mathsf C}u^{\mathsf c}\otimes\mathbb L_{-C}w_3^\gamma)\\
 &\qquad\qquad=(-1)^{\epsilon_1\epsilon_1'+\epsilon_2 \epsilon_3'}
 \rho_{\mathsf{F}}(\Psi_{-\mathsf A}\mathbf1,\Psi_{-\mathsf B}u^{\mathsf b},
                                      \Psi_{-\mathsf C}u^{\mathsf c})
 \rho_f^{(\sigma)}(\mathbf L_{-A}\phi_1,\mathbb L_{-B}w_2^\alpha,
                                      \mathbb L_{-C}w_3^\gamma),
\fe
where the fermion 3-point function is normalized by $\rho_{\mathsf{F}}(\mathbf{1,1,1})=1$, $\rho_{\mathsf{F}}(\mathbf{1},u^0,u^0)=1$ and $\rho_{\mathsf{F}}(\mathbf{1},u^1,u^1)=i$. 

As is, the enlarged block $\widehat{\mathbf{F}}^{(f,\sigma,a)}_{\Gamma, L}$ vanishes whenever the vertex signs $\sigma_v$ around a closed loop $C\subset \Gamma$ with all Ramond edges have product
\ie
\prod_{v\in C}\sigma_v = -1.
\fe
We call such a loop an odd Ramond loop, and the algorithm for computing superconformal blocks with odd Ramond loops will be presented in Section \ref{sec:oddloop}. In the rest of this section, we assume that there are no such odd Ramond loops in the labeled graph $(\Gamma, L)$.

The first way of computing the enlarged block is to take the basis $\Psi_{-\mathsf{A}}\mathbf{1}\otimes \mathbf{L}_{-A}\phi$ for states in $\mathcal{M}_\NS(h)$ and $\Psi_{-\mathsf{A}}u^{\mathsf{a}}\otimes \mathbb{L}_{-A}w^\alpha$ for $\mathcal{M}_{\R}(\beta)$. This allows us to express the enlarged block as a formal product of the superconformal block \eqref{scblock} and the free fermion block. Namely,
\ie \label{FconvF}
\widehat{\mathbf{F}}_{\Gamma, L}^{(f,\sigma,a)} = \mathbf{F}_{\Gamma, L}^{(f,\sigma,a)}\  \mathbf{F}^{\mathsf{F}}_{\Gamma, L},
\fe
where we view the blocks as formal series in both the plumbing parameters $q_e$ and the spin signs $\eta_e$, and the formal product acts on monomials as $(\prod_e q_e^{r_e}\eta_e^{\epsilon_e})(\prod_e q_e^{s_e}\eta_e^{\epsilon'_e})=\prod_e q_e^{r_e+s_e}\eta_e^{\epsilon_e+\epsilon'_e}$.
The free fermion block $\mathbf{F}^{\mathsf{F}}_{\Gamma,L}$ is defined as
\ie
\mathbf{F}^{\mathsf{F}}_{\Gamma, L} &= \sum_{\{\mathsf{A,a}\}} (-1)^{K_\Gamma(\epsilon')} \prod_{e\in E_\NS\cup E_\R} q_e^{|\mathsf{A}_{e_1}|}\eta_e^{\epsilon_e'}\prod_{e\in E_{\NS}}\mathbf{G}_{\F}^{\mathsf{A}_{e_1}\mathsf{A}_{e_2}}\prod_{e\in E_{\R}}\mathbb{G}_{\F}^{\mathsf{A}_{e_1}\mathsf{a}_{e_1},\mathsf{A}_{e_2}\mathsf{a}_{e_2}}\\
&\times \prod_{v\in V_\NS} \rho_{\mathsf{F}}(\Psi_{-\mathsf{A}_{v_1}}\mathbf{1},\Psi_{-\mathsf{A}_{v_2}}\mathbf{1},\Psi_{-\mathsf{A}_{v_3}}\mathbf{1})\prod_{v\in V_\R} \rho_{\mathsf{F}}(\Psi_{-\mathsf{A}_{v_1}}\mathbf{1},\Psi_{-\mathsf{A}_{v_2}}u^{\mathsf{a}_{v_2}},\Psi_{-\mathsf{A}_{v_3}}u^{\mathsf{a}_{v_3}}),
\fe
where the sum is subject to the same condition \eqref{Fselection}.

The enlarged block can also be computed using the basis $L_{-A}^{(1)}L_{-B}^{(2)}v_n$ for $\mathcal{M}_{\NS}(h)$ and $L_{-A}^{(1)}L_{-B}^{(2)}v_n^\alpha$ for $\mathcal{M}_{\R}(\beta)$. This leads to the following decomposition of $\widehat{\mathbf{F}}_{\Gamma, L}^{(f, \sigma, a)}$:
\ie
\widehat{\mathbf{F}}_{\Gamma,L}^{(f, \sigma, a)} &= \sum_{\{n,\alpha\}} (-1)^{K_{\Gamma}(\epsilon)} \prod_{e\in E_\NS} q_e^{2n_e^2}\eta_e^{\epsilon_e} \prod_{e\in E_\R} q_e^{2n_e^2-\frac{1}{8}}\eta_e^{\alpha_e}\\
&\times \prod_{i=1}^2 F_{\Gamma}(h_{e,n_e}^{(i)}, c^{(i)}) \prod_{v\in V_\NS} \mathbf{B}_{a_v}(n_{v_1},n_{v_2},n_{v_3})\\
&\times \prod_{v\in V_\R} \mathbb{B}_{f_v}^{(\sigma_v)}(n_{v_1},n_{v_2},n_{v_3},\alpha_{v_2},\alpha_{v_3}),\\
\fe
where the labels $n$ take values in $n\in \frac{1}{2}\mathbb{Z}$ for NS half-edges, and $n\in \frac{1}{2}\mathbb{Z}+\frac{1}{4}$ for R half-edges. $F_\Gamma(h_e; c)$ denotes the Virasoro conformal block in the same channel $\Gamma$ with intermediate weights $h_{e}$ and central charge $c$. The weights $h_{e,n_e}^{(i)}$ and central charges $c^{(i)}$ are given respectively by \eqref{h12} and \eqref{c12}. $\epsilon_e$ here denotes the Grassmann parity of the $\mathsf{Vir}\oplus \mathsf{Vir}$ primary $v_n$ or $v_n^\alpha$ that propagates on edge $e$. Explicitly, one has $\epsilon_e=2n_e + \epsilon(\phi_e)\text{ mod }2$ for NS sector and $\epsilon_e=\alpha_e$ for R sector. $\mathbf{B}_a, \mathbb{B}^{(\sigma)}_f$ are the branching coefficients, defined as
\ie \label{branch}
&\mathbf{B}_a(n_1, n_2, n_3) = \frac{\hat{\rho}_a(v_{1,n_1}, v_{2,n_2}, v_{3,n_3})}{\|v_{1,n_1}\|\|v_{2,n_2}\|\|v_{3,n_3}\|},\\
&\mathbb{B}^{(\sigma)}_f(n_1, n_2, n_3,\alpha_2, \alpha_3) = \frac{\hat{\rho}_f^{(\sigma)}(v_{1,n_1}, v_{2,n_2}^{\alpha_2}, v_{3,n_3}^{\alpha_3})}{\|v_{1,n_1}\|\|v_{2,n_2}^{\alpha_2}\|\|v_{3,n_3}^{\alpha_3}\|}.
\fe

The (NS, NS, NS) branching coefficients $\mathbf{B}_a$ have been worked out using the free field representation \cite{Hadasz:2013dza} and are given in \eqref{NSbranching}, while the (NS, R, R) coefficients have never been computed analytically in the literature. We now present a recursive algorithm for determining both numerically. It can be proven that, for $n> 0$, $L_1 v_n, L_{1}v_n^\alpha$ both only contain level-$(4n-3)$ states in the $\mathsf{Vir\oplus Vir}$ module generated by $v_{n-1}, v_{n-1}^\alpha$, while $L_{-1}v_n, L_{-1}v_n^\alpha$ only contain level-1 states in the $v_n, v_n^\alpha$ module as well as level-$(4n-1)$ states in the $v_{n-1}, v_{n-1}^\alpha$ module. Namely, there exist coefficients $\mathbf{V}_n^{(1,2)}, \mathbf{V}_{\pm 1, n}^{AB}, \mathbb{V}_{n,\alpha}^{(1,2)}, \mathbb{V}_{\pm 1, n,\alpha}^{AB}$ such that
\ie \label{Vcoeffs}
L_1 v_n^\alpha &= \sum_{|A|+|B|=4n-3} \mathbb{V}_{1,n,\alpha}^{AB} L_{-A}^{(1)}L_{-B}^{(2)} v_{n-1}^\alpha,\\
L_{-1} v_n^\alpha &= \mathbb{V}_{n,\alpha}^{(1)} L_{-1}^{(1)}v_n^\alpha + \mathbb{V}_{n,\alpha}^{(2)} L_{-1}^{(2)}v_n^\alpha + \sum_{|A|+|B|=4n-1} \mathbb{V}_{-1,n,\alpha}^{AB} L_{-A}^{(1)}L_{-B}^{(2)} v_{n-1}^\alpha,\\
L_1 v_n &= \sum_{|A|+|B|=4n-3} \mathbf{V}_{1,n}^{AB} L_{-A}^{(1)}L_{-B}^{(2)} v_{n-1},\\
L_{-1} v_n &= \mathbf{V}_{n}^{(1)} L_{-1}^{(1)}v_n + \mathbf{V}_{n}^{(2)} L_{-1}^{(2)}v_n  + \sum_{|A|+|B|=4n-1} \mathbf{V}_{-1,n}^{AB} L_{-A}^{(1)}L_{-B}^{(2)} v_{n-1}.
\fe
These coefficients $\mathbf{V}_n^{(1,2)}, \mathbf{V}_{\pm 1, n}^{AB}, \mathbb{V}_n^{(1,2)}, \mathbb{V}_{\pm 1, n}^{AB}$ can be determined directly by applying the free-field realization of $L_{\pm 1}$ on \eqref{vNS}, \eqref{vR}. A similar relation exists for $n< 0$, where $L_{\pm 1}v_n, L_{\pm 1}v_n^\alpha$ include states in the $v_{n+1}, v_{n+1}^\alpha$ module instead of $v_{n-1}, v_{n-1}^\alpha$.

After determining the coefficients $\mathbb{V}$ and $\mathbf{V}$, one can use the $L_1$ conformal Ward identities
\ie
\hat{\rho}_f^{(\sigma)}(L_1v_{1,n_1},v_{2,n_2}^{\alpha_2},v_{3,n_3}^{\alpha_3})&= \hat{\rho}_f^{(\sigma)}(v_{1,n_1},L_{-1}v_{2,n_2}^{\alpha_2},v_{3,n_3}^{\alpha_3})+\hat{\rho}_f^{(\sigma)}(v_{1,n_1},v_{2,n_2}^{\alpha_2},L_{-1}v_{3,n_3}^{\alpha_3}),\\
\hat{\rho}_f^{(\sigma)}(v_{1,n_1},L_1v_{2,n_2}^{\alpha_2},v_{3,n_3}^{\alpha_3})&= \hat{\rho}_f^{(\sigma)}(L_{-1}v_{1,n_1},v_{2,n_2}^{\alpha_2},v_{3,n_3}^{\alpha_3})-\hat{\rho}_f^{(\sigma)}(v_{1,n_1},L_{-1}v_{2,n_2}^{\alpha_2},v_{3,n_3}^{\alpha_3})\\
&\quad-2\hat{\rho}_f^{(\sigma)}(v_{1,n_1},L_0v_{2,n_2}^{\alpha_2},v_{3,n_3}^{\alpha_3})-\hat{\rho}_f^{(\sigma)}(v_{1,n_1},v_{2,n_2}^{\alpha_2},L_1v_{3,n_3}^{\alpha_3}),
\fe
which become recursion relations for the branching coefficients $\mathbb{B}(n_1, n_2, n_3)$. Given the boundary values $\mathbb{B}(n_1, n_2, n_3)$ with $-\frac{1}{2}\leq n_i \leq \frac{1}{2}$, which can be computed analytically using the free field realization, one can solve for $\mathbb{B}(n_1, n_2, n_3)$ recursively. For example, for $n_1\geq 1$ and $n_2, n_3\geq \frac{3}{4}$, the first equation becomes
\ie
\Bigg[\sum_{i=1}^2 &(\mathbb{V}_{n_2,\alpha_2}^{(i)}-\mathbb{V}_{n_3,\alpha_3}^{(i)})(h_{2,n_2}^{(i)}+h_{3,n_3}^{(i)}-h_{1,n_1}^{(i)})\Bigg]\mathbb{B}_f^{(\sigma)}(n_1,n_2,n_3,\alpha_2,\alpha_3)=\\
& - \sum_{|A|+|B|=4n_1-3}\mathbf{V}_{1,n_1}^{AB}\mathbb{B}_f^{(\sigma)}(n_1-1,n_2,n_3,\alpha_2,\alpha_3)\frac{\|v_{n_1-1}\|}{\|v_{n_1}\|}\\
& \qquad \qquad \qquad \qquad \times \rho(L_{-A}\nu_{h_{1,n_1-1}^{(1)}}, \nu_{h_{2,n_2}^{(1)}}, \nu_{h_{3,n_3}^{(1)}})\rho(L_{-B}\nu_{h_{1,n_1-1}^{(2)}}, \nu_{h_{2,n_2}^{(2)}}, \nu_{h_{3,n_3}^{(2)}})\\
& + \sum_{|C|+|D|=4n_2-1} \mathbb{V}_{-1,n_2, \alpha_2}^{CD}\mathbb{B}_f^{(\sigma)}(n_1,n_2-1,n_3,\alpha_2,\alpha_3)\frac{\|v_{n_2-1}^{\alpha_2}\|}{\|v_{n_2}^{\alpha_2}\|}\\
& \qquad \qquad \qquad \qquad \times \rho(\nu_{h_{1,n_1}^{(1)}},L_{-C}\nu_{h_{2,n_2-1}^{(1)}}, \nu_{h_{3,n_3}^{(1)}})\rho(\nu_{h_{1,n_1}^{(2)}},L_{-D}\nu_{h_{2,n_2-1}^{(2)}}, \nu_{h_{3,n_3}^{(2)}})\\
& + \sum_{|E|+|F|=4n_3-1} \mathbb{V}_{-1,n_3, \alpha_3}^{EF}\mathbb{B}_f^{(\sigma)}(n_1,n_2,n_3-1,\alpha_2,\alpha_3)\frac{\|v_{n_3-1}^{\alpha_3}\|}{\|v_{n_3}^{\alpha_3}\|}\\
& \qquad \qquad \qquad \qquad \times \rho(\nu_{h_{1,n_1}^{(1)}},\nu_{h_{2,n_2}^{(1)}}, L_{-E}\nu_{h_{3,n_3-1}^{(1)}})\rho(\nu_{h_{1,n_1}^{(2)}},\nu_{h_{2,n_2}^{(2)}}, L_{-F}\nu_{h_{3,n_3-1}^{(2)}}),
\fe
which relates the branching coefficient $\mathbb{B}^{(\sigma)}_f$ with those of less total $n$. These low-level branching coefficients, together with the ratio of BPZ norms of $\mathsf{Vir}\oplus \mathsf{Vir}$ primaries, can all be computed using the free field representation, whose results are reviewed in Appendix \ref{app:freefield}.

Once the branching coefficients are solved, one can use the $c$-recursion relation to compute the Virasoro conformal blocks \cite{Cho:2017oxl}, which computes $\widehat{\mathbf{F}}_{\Gamma,L}^{(f, \sigma, a)}$ as a formal series in the plumbing parameters and the spin lift signs. The free fermion block $\mathbf{F}^{\mathsf{F}}_{\Gamma,L}$ can be computed from its definition efficiently due to the small size of the space of fermion descendants. The superconformal block $\mathbf{F}_{\Gamma,L}^{(f, \sigma, a)}$, as a series in the plumbing parameters, can be extracted coefficient by coefficient from the product \eqref{FconvF}.

\section{Blocks with odd Ramond loops}

\label{sec:oddloop}

\begin{figure}
\centering
\definecolor{splitpink}{RGB}{255,105,166}
\begin{tikzpicture}[
  x=1cm, y=1cm,
  line cap=round, line join=round,
  sphere/.style={line width=0.65pt},
  equator/.style={sphere, dash pattern=on 6pt off 6pt},
  old tube/.style={draw=black!40, line width=1.1pt, dashed},
  new tube/.style={draw=splitpink, line width=0.6pt, dashed},
  every node/.style={inner sep=2pt, font=\fontsize{9}{11}\selectfont},
  graph label/.style={font=\fontsize{14}{17}\selectfont}
]

\draw[line width=1.2pt] (2.08,0) circle[radius=2.08];
\draw[splitpink,line width=1.2pt] (0,0) -- (2.08,0) -- (4.16,0);
\draw[splitpink,line width=1.2pt] (2.08,0) -- (2.62,0.97);
\foreach \x in {0,2.08,4.16}
  \fill (\x,0) circle[radius=2.1pt];
\node[graph label,above] at (2.08,2.20) {$q_1$};
\node[graph label,above] at (1.08,0.13) {$q_L$};
\node[graph label,above] at (3.16,0.13) {$q_R$};
\node[graph label]       at (2.08,-1.76) {$q_3$};

\foreach \x/\col in {7.10/black,10.89/splitpink,14.68/black} {
  \draw[sphere,\col] (\x,0.07) circle[radius=1.41];
  \draw[equator,\col] (\x,0.07) ellipse[x radius=1.41,y radius=0.28];
}

\coordinate (Linf)  at (7.10, 1.39);
\coordinate (Lone)  at (7.88,-0.16);
\coordinate (Lzero) at (7.10,-1.26);
\coordinate (Rinf)  at (14.68,1.40);
\coordinate (Rone)  at (15.46,-0.14);
\coordinate (Rzero) at (14.68,-1.26);

\coordinate (Winf)  at (10.89,1.36);
\coordinate (Wone)  at (11.69,-0.16);
\coordinate (Wzero) at (10.94,-1.26);

\draw[old tube] (Linf) .. controls (9.63,2.74) and (12.15,2.74) .. (Rinf);
\draw[old tube] (Linf) .. controls (9.63,2.65) and (12.15,2.65) .. (Rinf);
\draw[old tube] (Lzero) .. controls (9.63,-2.48) and (12.15,-2.48) .. (Rzero);
\draw[old tube] (Lzero) .. controls (9.63,-2.39) and (12.15,-2.39) .. (Rzero);

\draw[new tube] (Lone) .. controls (8.85,-0.78) and (10.02,-1.18) .. (Wzero);
\draw[new tube] (Lone) .. controls (8.85,-0.84) and (10.02,-1.24) .. (Wzero);

\draw[new tube] (Winf) .. controls (12.45,1.62) and (14.20,0.99) .. (Rone);
\draw[new tube] (Winf) .. controls (12.45,1.56) and (14.20,0.93) .. (Rone);

\foreach \p in {Linf,Lone,Lzero,Rinf,Rone,Rzero}
  \fill (\p) circle[radius=1.1pt];
\foreach \p in {Winf,Wone,Wzero}
  \fill[splitpink] (\p) circle[radius=1.1pt];

\node[above left]  at (6.89,1.48) {$z_1=\infty$};
\node[above right] at (14.73,1.48) {$z_2=\infty$};
\node[above left]  at (7.87,-0.07) {$z_1=1$};
\node[below left]  at (15.57,-0.35) {$z_2=1$};
\node[below]       at (7.10,-1.48) {$z_1=0$};
\node[below]       at (14.68,-1.48) {$z_2=0$};
\node[above]       at (10.89,1.52) {$y=\infty$};
\node[above left]  at (11.58,-0.07) {$y=1$};
\node[below]       at (10.89,-1.48) {$y=0$};

\node[above] at (10.89,2.51) {$q_1$};
\node[below] at (10.89,-2.27) {$q_3$};
\node[fill=none,inner sep=1.5pt] at (9.25,-0.50) {$q_L$};
\node[fill=none] at (13.17,1.52) {$q_R$};

\end{tikzpicture}
    \caption{Splitting the second edge in the genus two theta channel. The second edge that connects $z_1=1, z_2=1$ in the original plumbing graph is replaced by a new sphere and two new edges shown in pink. The tubes implement the plumbing identifications $\frac{1}{z_1z_2}=q_1$, $(z_1-1)y=q_L$, $\frac{z_2-1}{y}=q_R$, and $z_1z_2=q_3$.}
    \label{fig:splitedge}
\end{figure}
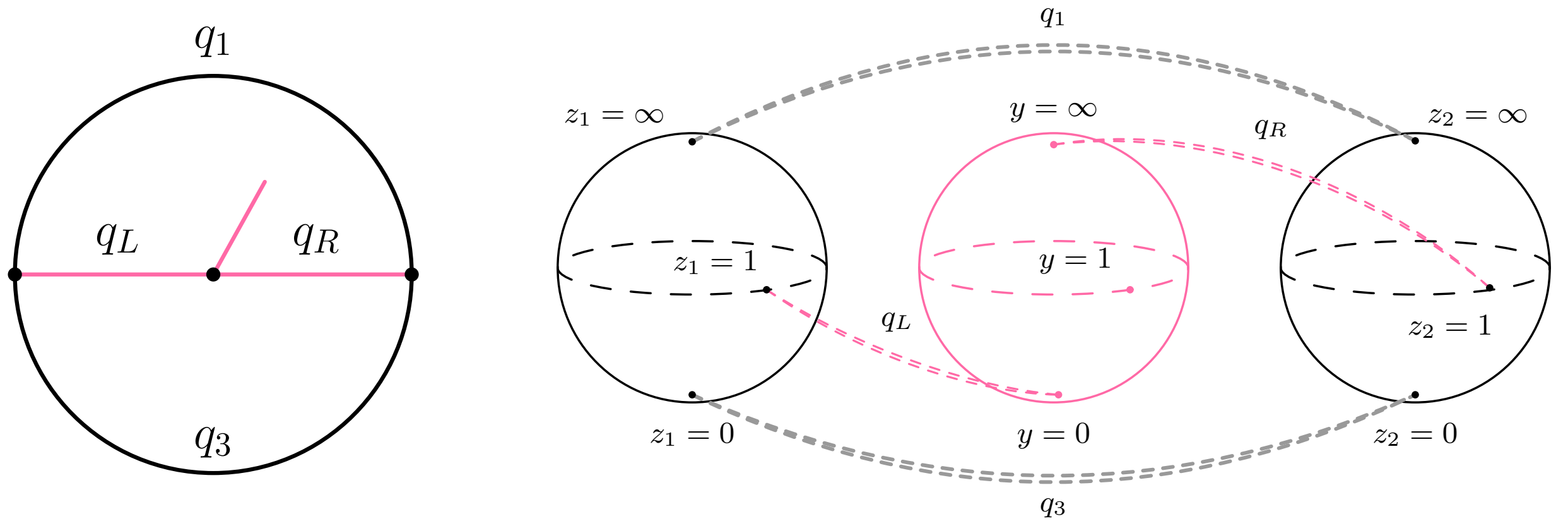

We now turn to the case where the labeled graph $(\Gamma, L)$ contains odd Ramond loops, where the enlarged block $\widehat{\mathbf{F}}_{\Gamma,L}^{(f,\sigma,a)}$ vanishes because the intermediate states built from the fermionic ground state $u^0$ along the loop give contributions opposite to those from $u^1$.

To remove this cancellation, we define a new enlarged block $\widehat{\mathbf{F}}^{(f,\sigma,a)}_{\Gamma', L'}[\Theta\psi]$ by essentially inserting a free fermion operator $\psi$ onto the odd Ramond loop. For each odd Ramond loop $C$, we split one of the edges $e$ in $C$. Suppose $e$ connects two Riemann spheres with complex coordinates $z, z'$, and the punctures $e$ connects to are located at $z=z_0, z'=z_0'$. We replace $e$ by introducing a new Riemann sphere with complex coordinate $y$, and glue the two punctures $y=0, y=\infty$ with the two original punctures $z_0, z_0'$ by
\ie 
(z-z_0)y={q}_{e_L},\quad (z'-z'_0)y^{-1}=q_{e_R}.
\fe
The two new edges $e_L, e_R$ formed are both Ramond. We repeat this splitting procedure for each odd Ramond loop in the labeled graph $(\Gamma, L)$, and we denote the sets of all split and unsplit Ramond edges in the original graph $\Gamma$ by $E_\mathrm{split}, E_{\mathrm{unsplit}}$, respectively. This procedure defines a plumbing channel corresponding to a new labeled graph $(\Gamma', L')$, whose plumbing parameters are given by $\{q_e\}_{e\notin E_\mathrm{split}}$ and $\{q_{e_L}, q_{e_R}\}_{e\in E_\mathrm{split}}$. The original plumbing parameter $q_e$ for a split edge can be related to $q_{e_L}, q_{e_R}$ by $q_e=q_{e_L}q_{e_R}$. Similarly, the spin lift sign $\eta_e$ can be recovered by $\eta_e=\eta_{e_L}\eta_{e_R}$. An example of this procedure where one splits the second edge of the genus two theta channel is illustrated in figure \ref{fig:splitedge}. 

On each new sphere $y$, we insert a free fermion operator $\psi$ at $y=1$, together with a ground state reflection operator $\Theta$ that acts on the fermion ground states as
\ie
\Theta u^{\mathsf{b}} = (-1)^{\mathsf{b}+1}u^{1-\mathsf{b}},
\fe
and it anticommutes with $\psi_r, G_r$ while commuting with $L_n$. This ground state reflection operator is a topological line operator, and we insert it on a circle with $|y|>1$. The new enlarged block $\widehat{\mathbf{F}}^{(f,\sigma,a)}_{\Gamma', L'}[\Theta\psi]$ is defined to be
\ie 
\widehat{\mathbf{F}}^{(f,\sigma,a)}_{\Gamma', L'}[\Theta\psi] &= \sum_{\{A,\alpha,\mathsf{A},\mathsf{a}\}} (-1)^{K_{\Gamma'}(\epsilon+\epsilon')} \prod_{e\in E_\NS\cup E_{\mathrm{unsplit}}} q_e^{|A_{e_1}|+|\mathsf{A}_{e_1}|}\eta_e^{\epsilon_e+\epsilon_e'} (-1)^{\epsilon_e\epsilon_e'}\\
&\times \prod_{e\in E_\NS} \mathbf{G}_{\F}^{\mathsf{A}_{e_1}\mathsf{A}_{e_2}}  \mathbf{G}_{h_e}^{A_{e_1}A_{e_2}} \prod_{e\in E_{\mathrm{unsplit}}} \mathbb{G}_{\F}^{\mathsf{A}_{e_1}\mathsf{a}_{e_1},\mathsf{A}_{e_2}\mathsf{a}_{e_2}} \mathbb{G}_{\beta_e}^{A_{e_1}\alpha_{e_1},A_{e_2}\alpha_{e_2}}\\
&\times \prod_{v \in V_\NS} \hat{\rho}_{a_v}(\Psi_{-\mathsf{A}_{v_1}}\mathbf{1}\otimes \mathbf L_{-A_{v_1}}\phi_{v_1},\Psi_{-\mathsf{A}_{v_2}}\mathbf{1}\otimes \mathbf L_{-A_{v_2}}\phi_{v_2},\Psi_{-\mathsf{A}_{v_3}}\mathbf{1}\otimes \mathbf L_{-A_{v_3}}\phi_{v_3})\\
&\times\prod_{v\in V_\R}\hat{\rho}_{f_v}^{(\sigma_v)}(\Psi_{-\mathsf{A}_{v_1}}\mathbf{1}\otimes \mathbf L_{-A_{v_1}}\phi_{v_1},\Psi_{-\mathsf{A}_{v_2}} u^{\mathsf{a_{v_2}}}\otimes\mathbb L_{-A_{v_2}}w_{v_2}^{\alpha_{v_2}},\Psi_{-\mathsf{A}_{v_3}} u^{\mathsf{a_{v_3}}}\otimes\mathbb L_{-A_{v_3}}w_{v_3}^{\alpha_{v_3}})\\
&\times \prod_{e\in E_\mathrm{split}} q_{e_L}^{|A_{e_1}|+|\mathsf{A}_{e_1}|}q_{e_R}^{|A_{e_2}|+|\mathsf{A}_{e_2}|}(\eta_{e_L}\eta_{e_R})^{\epsilon_e+\epsilon_e'}\\
&\qquad \qquad \times \mathbb{G}_{\F}^{\mathsf{A}_{e_1}\mathsf{a}_{e_1},\mathsf{A}_{e_L}\mathsf{a}_{e_L}} \mathbb{G}_{\beta_e}^{A_{e_1}\alpha_{e_1},A_{e_L}\alpha_{e_L}}\mathbb{G}_{\F}^{\mathsf{A}_{e_R}\mathsf{a}_{e_R},\mathsf{A}_{e_2}\mathsf{a}_{e_2}} \mathbb{G}_{\beta_e}^{A_{e_R}\alpha_{e_R},A_{e_2}\alpha_{e_2}}\\
&\qquad \qquad \times \langle \langle \Psi_{-\mathsf{A}_{e_2}}u^{\mathsf{a}_{e_2}}\otimes \mathbb{L}_{-A_{e_R}}w^{\alpha_{e_R}}_{e_R}|\Theta \psi(1)|\Psi_{-\mathsf{A}_{e_1}}u^{\mathsf{a}_{e_1}}\otimes \mathbb{L}_{-A_{e_L}}w^{\alpha_{e_L}}_{e_L}\rangle,
\fe
where the sum is over all displayed labels, subject to \eqref{selection-block} and \eqref{Fselection}, except that the fermion labels $\mathsf{A}_{e_1}$, $\mathsf{A}_{e_2}$ and $\mathsf{a}_{e_1}$, $\mathsf{a}_{e_2}$ on a split edge $e\in E_{\mathrm{split}}$ may differ due to the $\Theta\psi(1)$ insertion. Their combined parities are constrained to be the same, namely
\ie
(-1)^{\mathsf{A}_{e_1}+\mathsf{a}_{e_1}}=(-1)^{\mathsf{A}_{e_2}+\mathsf{a}_{e_2}},\quad e\in E_\mathrm{split}.
\fe

This enlarged block is non-vanishing, and can be expressed as a product of the ordinary superconformal block and the free fermion block $\mathbf{F}^{\mathsf{F}}_{\Gamma', L'}[\Theta \psi]$ with the same insertion $\Theta \psi(1)$. Namely,
\ie \label{Insconv}
\widehat{\mathbf{F}}^{(f,\sigma,a)}_{\Gamma', L'}[\Theta\psi] = {\mathbf{F}}^{(f,\sigma,a)}_{\Gamma, L} \mathbf{F}^{\mathsf{F}}_{\Gamma', L'}[\Theta \psi],
\fe
where one restricts to the terms in $\widehat{\mathbf{F}}^{(f,\sigma,a)}_{\Gamma', L'}[\Theta\psi]$ and $\mathbf{F}^{\mathsf{F}}_{\Gamma', L'}[\Theta \psi]$ that have the same powers of $q_{e_L}, q_{e_R}$ and $\eta_{e_L}, \eta_{e_R}$ for each split edge $e$, and replaces $q_{e_L}^m q_{e_R}^m$ by $q_e^m$, $\eta_{e_L}^\epsilon \eta_{e_R}^\epsilon$ by $\eta_{e}^\epsilon$ before performing the convolution. It is evident that the superconformal block extracted from this convolution only depends on the product $q_{e}=q_{e_L}q_{e_R}$, but not the individual parameters $q_{e_L},q_{e_R}$ of the split edges.

On the other hand, since the operator $\psi$ is proportional to the $\mathsf{Vir\oplus Vir}$ primary $v_{1/2}(P)$ with Liouville momentum $P=\frac{Q}{2}$, and the operator $\Theta$ commutes with the two Virasoro subalgebras, one can decompose $\widehat{\mathbf{F}}^{(f,\sigma,a)}_{\Gamma', L'}[\Theta \psi]$ into Virasoro blocks as
 \ie
\widehat{\mathbf{F}}^{(f,\sigma,a)}_{\Gamma', L'}[\Theta \psi]&= \sum_{\{n,\alpha\}} (-1)^{K_{{\Gamma'}}(\epsilon)} \prod_{e\in E_\NS} q_e^{2n_e^2}\eta_e^{\epsilon_e} \prod_{e\in E_\R\backslash E_\mathrm{split}} q_e^{2n_e^2-\frac{1}{8}}\eta_e^{\alpha_e}\\
&\times \prod_{e\in E_\mathrm{split}} q_{e_L}^{2n_{e_L}^2-\frac{1}{8}}q_{e_R}^{2n_{e_R}^2-\frac{1}{8}} \eta_e^{\alpha_e}\ \mathbb{S}(n_{e_R}, n_{e_L}, \alpha_e)\\
&\times \prod_{i=1}^2 F_{\Gamma'}(h_{e, n_e}^{(i)},h_\psi^{(i)}, c^{(i)}) \prod_{v\in V_\NS} \mathbf{B}_{a_v}(n_{v_1},n_{v_2},n_{v_3}) \prod_{v\in V_\R} \mathbb{B}_{f_v}^{(\sigma_v)}(n_{v_1},n_{v_2},n_{v_3},\alpha_{v_2},\alpha_{v_3}),
\fe   
where $F_{\Gamma'}(h_{e}^{(i)},h_\psi^{(i)}, c^{(i)})$ stands for the Virasoro conformal block in the new channel $\Gamma'$ with a primary with external conformal weights
\ie 
h_\psi^{(1)}=-\frac{1+2 b^2}{2\left(1-b^2\right)}, \quad h_\psi^{(2)}=\frac{b^2+2}{2\left(1-b^2\right)}
\fe
inserted on the punctures originally with $\psi(1)$ insertions. The splitting coefficient $\mathbb{S}(n_R,n_L,\alpha)$ is defined as
\ie
\mathbb{S}(n_R, n_L,\alpha) = \frac{\langle \langle v_{n_R}^\alpha|\Theta \psi(1)|v_{n_L}^{\alpha}\rangle}{\|v_{n_L}^{\alpha}\|\|v_{n_R}^{\alpha}\|},
\fe
which is nonvanishing only for $n_R = n_L \pm \frac{1}{2}$.

Once the coefficients $\mathbb{V}_{n,\alpha}^{1,2}, \mathbb{V}_{\pm 1,n,\alpha}^{AB}$ in \eqref{Vcoeffs} are known, the splitting coefficients can be solved for using the $L_1$ conformal Ward identity 
\ie
\langle\langle L_1 v_{n_R}^\alpha| \Theta \psi(1)|v_{n_L}^\alpha\rangle=\langle\langle v_{n_R}^\alpha| \Theta \psi(1)|L_{-1}v_{n_L}^\alpha\rangle,
\fe
which becomes a recursion relation for the splitting coefficients after plugging in \eqref{Vcoeffs}. After solving for the splitting coefficients and the branching coefficients, one can compute $\widehat{\mathbf{F}}^{\{f,\sigma,a\}}_{\Gamma', L'}[\Theta \psi]$ using this decomposition, and the physical superconformal block $\mathbf{F}^{\{f,\sigma,a\}}_{\Gamma, L}$ can be extracted from the product \eqref{Insconv}.

\section{Numerical Tests}

\label{sec:numerics}

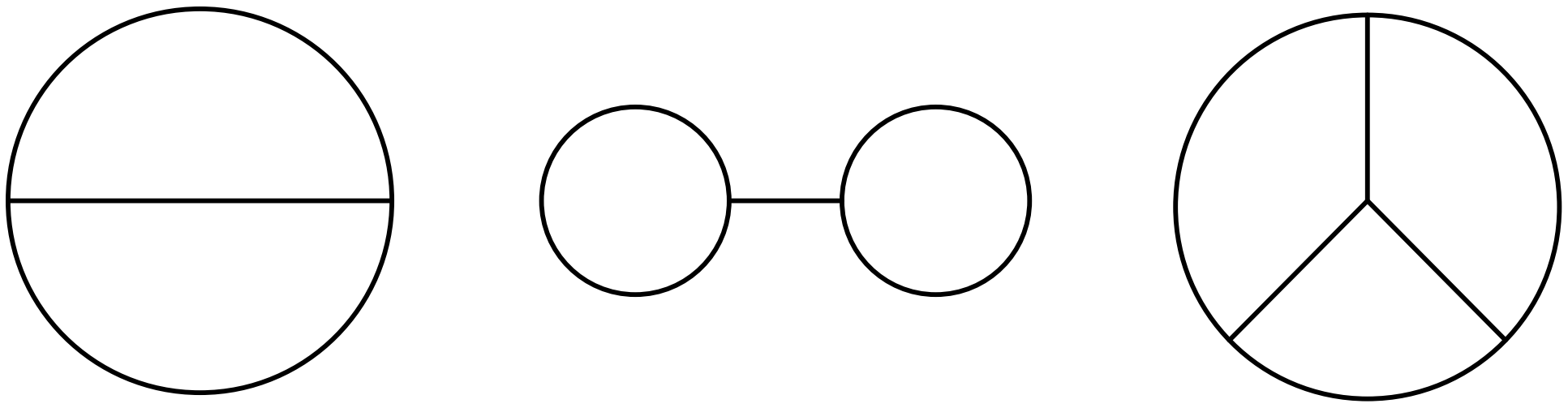
\begin{figure}
\begin{center}
\begin{tikzpicture}[
  x=1cm, y=1cm,
  draw=black, line width=1.1pt,
  line cap=round, line join=round,
  every node/.style={fill=none}
]

\begin{scope}[shift={(1.565,0)}]
  \draw (0,0) circle[radius=1.565];
  \draw (-1.565,0) -- (1.565,0);
\end{scope}

\begin{scope}[shift={(5.115,0)}]
  \draw (0,0) circle[radius=0.765];
  \draw (2.45,0) circle[radius=0.765];
  \draw (0.765,0) -- (1.685,0);
\end{scope}

\begin{scope}[shift={(11.085,-0.05)}]
  \draw (0,0) circle[radius=1.565];
  \draw (90:1.565) -- (0,0.05);
  \draw (224:1.565) -- (0,0.05) -- (316:1.565);
\end{scope}

\end{tikzpicture}
\end{center}
    \caption{From left to right: The genus two theta channel, the genus two glasses channel, and the genus three tetrahedron channel.}
    \label{fig:chanel}
\end{figure}

Using the algorithm presented in Sections \ref{sec:blockdecomp} and \ref{sec:oddloop}, one can compute arbitrary superconformal blocks as a series in the plumbing parameters. We have tested our algorithm in the genus two theta and glasses channels, as well as the genus three tetrahedron channel. The plumbing graphs of these channels are shown in figure \ref{fig:chanel}.

On a laptop and working with a precision of 40 digits, computing the genus two theta channel block with NS-R-R edges takes approximately 6 minutes up to level $n_1, n_2, n_3\leq 10$ in the case without an odd Ramond loop, and takes 26 minutes for the case with an odd Ramond loop.\footnote{These timings were measured using a single core of an Apple M2 processor.} A direct computation using the definition \eqref{scblock} is estimated to take more than 6 hours for the same block on the same device with the same precision.

For all the spin structures in the two genus two channels, the low-level coefficients of $q_1^{n_1}q_2^{n_2}q_3^{n_3}$ in $\mathbf{F}_{\Gamma, L}^{(f,\sigma, a)}$ obtained from our algorithm agree with direct computations from the definition \eqref{scblock} up to level $n_1, n_2, n_3\leq 5$ for all combinations of $f_v, \sigma_v, a_v$. The same test has been performed for the genus three tetrahedron channel, where the edges around one face are Ramond while the other edges are NS. For this channel, the block $\mathbf{F}_{\Gamma, L}^{(f,\sigma, a)}$ agrees with its definition up to total level $\sum_{i=1}^6 n_i \leq 5$. The superconformal block computed from our algorithm also agrees with the result from $c$-recursion \cite{Belavin:2018hfm} up to level $n_1, n_2, n_3 \leq 10$ for the genus two theta channel and up to total level $\sum_{i=1}^6 n_i \leq 10$ for the genus three tetrahedron channel, both with all NS internal edges.

Moreover, there are pairs of plumbing data $\{q_e^{\NS\NS\NS}, \eta_e^{\NS\NS\NS}\}$ and $\{q_e^{\NS\R\R}, \eta_e^{\NS\R\R}\}$ in the all-NS theta channel and the NS-R-R theta channel that parametrize the same spin Riemann surface, and we have checked that the following equation holds:
\ie 
\frac{Z_\mathrm{\mathrm{SL}}(q_e^{\NS\NS\NS}, \eta_e^{\NS\NS\NS})}{Z_\mathrm{\mathrm{SL}}(q_e^{\NS\R\R}, \eta_e^{\NS\R\R})} = \left(\frac{Z_{\mathrm{free}}(q_e^{\NS\NS\NS}, \eta_e^{\NS\NS\NS})}{Z_{\mathrm{free}}(q_e^{\NS\R\R}, \eta_e^{\NS\R\R})}\right)^{1+2Q^2},
\fe
where $Z_{\mathrm{SL}}$ stands for the partition function of the super-Liouville theory, and $Z_{\mathrm{free}}$ stands for the free boson + fermion partition function. The details of this check are explained in Appendix \ref{app:Zratiocheck}. The codes that implement our algorithm and perform the mentioned cross-checks are available in the GitHub repository.

\section{Discussion}

\label{sec:disc}

In this work, we develop the algorithm for numerically computing superconformal blocks at arbitrary genus, in arbitrary plumbing channels, for arbitrary spin structures. The double Virasoro subalgebra allows us to decompose the enlarged $\mathsf{SCA}_\nu \oplus \mathsf{F}_\nu$ block into Virasoro conformal blocks, with the coefficients given by branching coefficients, which are known on an (NS, NS, NS) sphere and can be solved numerically using a recursion relation on an (NS, R, R) sphere.

Our algorithm provides new tools for higher-loop superstring amplitudes. For two dimensional Type 0B and Type 0A string theory, such amplitudes can be used to further test the duality with matrix quantum mechanics, as well as the T-duality between the two theories \cite{Takayanagi:2003sm,Douglas:2003up,Balthazar:2022atu,Balthazar:2022apu}. They also permit direct worldsheet tests of matrix-integral predictions for the recently proposed super Virasoro minimal strings \cite{Johnson:2024fkm,Johnson:2025vyz,Johnson:2026ugu,Rangamani:2025wfa,Eberhardt:2026hfh}. One particularly interesting direction is the two dimensional $SO(23)$ and $SO(7)\times E_8$ heterotic string \cite{Davis:2005qe}, where the matrix quantum mechanical dual remains unknown, and the higher-loop or higher-point amplitudes could be used to help pin it down. In these applications of string amplitudes, our algorithm can be used to compute the integrand on moduli space. The integration can be performed using a similar method to \cite{Christian:new}: For genus $g\leq 3$, one samples the moduli space parametrized by the period matrix and converts the period matrix into plumbing parameters using a similar algorithm to \cite{Christian:2026qqy}. 

Beyond string amplitudes, modular covariance of higher-genus partition functions can be used to constrain the spectrum and the structure constants of $\mathcal{N}=1$ SCFTs, extending the analysis of \cite{Cho:2017fzo}. The superconformal blocks themselves are conjectured to be the wavefunctions of 3D  supergravity in AdS${}_3$ \cite{Eberhardt:2026hfh}, and the numerical values of the blocks obtained from our algorithm can be used to compute gravitational path integrals in this theory \cite{Bhattacharyya:2024vnw}.

\section*{Acknowledgments}

We are grateful to Ben Mazel, Sounak Pal, Victor A. Rodriguez, and Jaroslav Scheinpflug for discussions. This work is supported by DOE grant DE-SC0007870. YW and XY thank the RIKEN Interdisciplinary Theoretical and Mathematical Sciences Program (iTHEMS) and the Yukawa Institute for Theoretical Physics (YITP) at Kyoto University, where this work was completed during the iTHEMS-YITP workshop `Resurgence, Matrix model, String field theory and Instantons' (YITP-W-26-11). ChatGPT 5.6 Sol and 6 Astra were used in this work for drafting and editing, literature searches, algorithm implementation, and conducting numerical tests. 

\begin{appendix}

\section{Conventions}

\label{app:conventions}

In this appendix, we summarize the choice of conventions used in this paper. 

\paragraph{The superconformal algebra}
The $\mathcal{N}=1$ superconformal algebra $\mathsf{SCA}_\nu$ is given by
\ie
\left[L_m,L_n\right]
&=(m-n)L_{m+n}
+\frac{c}{12}\left(m^3-m\right)\delta_{m+n,0},\\
\left[L_m,G_r\right]
&=\left(\frac{m}{2}-r\right)G_{m+r},\\
\left\{G_r,G_s\right\}
&=2L_{r+s}
+\frac{c}{3}\left(r^2-\frac14\right)\delta_{r+s,0},
\fe
where $m,n\in \mathbb{Z}$ and $r,s\in \mathbb{Z}+\nu$, with $\nu = 0$ for the R sector and $\frac12$ for the NS sector.

\paragraph{BPZ conjugation}

The BPZ conjugation is defined using the inversion map $I: z\mapsto w=1/z$, and we choose the spin lift of this map such that
\ie
\left(\frac{dw}{dz}\right)^{\frac12} = -\frac{i}{z}.
\fe
On each plumbed edge with plumbing map $zw=q$, we choose the spin lift of the plumbing map such that 
\ie
\left(\frac{dw}{dz}\right)^{\frac12} = -\frac{i\sqrt{q}}{z},
\fe
with the branch of $\sqrt{q}$ chosen such that $\sqrt{q}$ for positive real $q$ gives the positive square root. Under BPZ conjugation, the superconformal generators and the auxiliary fermion transform as
\ie \label{BPZgenerator}
L_n^\dagger = L_{-n},\quad G_r^\dagger = iG_{-r},\quad \psi_r^\dagger = -i\psi_{-r}.
\fe

We also adopt the convention that the BPZ conjugation acts on products as 
\ie 
(XY)^\dagger = (-1)^{\epsilon_X\epsilon_Y}Y^\dagger X^\dagger,
\fe
where $\epsilon_X, \epsilon_Y$ are the Grassmann parities of $X, Y$. For tensor product operators, the BPZ inner product is given by
\ie \label{BPZtensor}
\langle \langle A\otimes X| B\otimes Y\rangle =(-1)^{\epsilon_B\epsilon_X}\langle \langle A|B\rangle\langle \langle X|Y\rangle.
\fe

\paragraph{Ward identities}

The $L_n$ Ward identity is given by
\ie
&\sum_{j\ge0}(-1)^j\binom nj\rho(L_{j-m-n+1}x_1,x_2,x_3)\\
&\quad=\sum_{j\ge0}\binom mj\rho(x_1,L_{n+j-1}x_2,x_3)
   +(-1)^n\sum_{j\ge0}(-1)^j\binom nj\rho(x_1,x_2,L_{m+j-1}x_3).
\fe
Here, $\rho$ is either $\rho_a$ or $\rho_f^{(\sigma)}$. The $G_r$ and $\psi_r$ Ward identities require one to choose a branch for the square root $\sqrt{z(z-1)}$, which we choose to be the following
\ie
\sqrt{z(z-1)}=
 \begin{cases}
 z\sqrt{1-z^{-1}},&z\sim\infty,\\
 \sqrt{z-1}\sqrt{1+(z-1)},&z\sim1,\\
 i\sqrt z\sqrt{1-z},&z\sim0.
 \end{cases}
\fe
With this choice of branch, the $G_r$ and $\psi_r$ Ward identities are given by
\ie
&\sum_{j\ge0}(-1)^j\binom nj
       \rho_a(G_{j-m-n+\frac12}x_1,x_2,x_3)\\
 &=i(-1)^{\epsilon_1}\sum_{j\ge0}\binom mj
       \rho_a(x_1,G_{n+j-\frac12}x_2,x_3)
   +i(-1)^{\epsilon_1+\epsilon_2+n}
      \sum_{j\ge0}(-1)^j\binom nj
                      \rho_a(x_1,x_2,G_{m+j-\frac12}x_3),
\fe \ie
 &\sum_{j\ge0}(-1)^j\binom nj
       \rho_{\F}(\psi_{j-m-n-\frac12}x_1,x_2,x_3)\\
 &=-i(-1)^{\epsilon'_1}\sum_{j\ge0}\binom mj
       \rho_{\F}(x_1,\psi_{n+j+\frac12}x_2,x_3)
   -i(-1)^{\epsilon'_1+\epsilon'_2+n}
      \sum_{j\ge0}(-1)^j\binom nj
                      \rho_{\F}(x_1,x_2,\psi_{m+j+\frac12}x_3).
\fe \ie
&\sum_{j\ge0}(-1)^j\binom{n+\frac12}{j}
 \rho_f^{(\sigma)}(G_{j-m-n-\frac12}x_1,x_2,x_3)\\
 &=i(-1)^{\epsilon_1}\sum_{j\ge0}\binom{m+\frac12}{j}
 \rho_f^{(\sigma)}(x_1,G_{n+j}x_2,x_3)
 -(-1)^{\epsilon_1+\epsilon_2+n}
       \sum_{j\ge0}(-1)^j\binom{n+\frac12}{j}
 \rho_f^{(\sigma)}(x_1,x_2,G_{m+j}x_3),
\fe \ie
 &\sum_{j\ge0}(-1)^j\binom{n-\frac12}{j}
 \rho_{\F}(\psi_{j-m-n+\frac12}x_1,x_2,x_3)\\
 &=-i(-1)^{\epsilon'_1}\sum_{j\ge0}\binom{m-\frac12}{j}
 \rho_{\F}(x_1,\psi_{n+j}x_2,x_3)
 -(-1)^{\epsilon'_1+\epsilon'_2+n}
       \sum_{j\ge0}(-1)^j\binom{n-\frac12}{j}
 \rho_{\F}(x_1,x_2,\psi_{m+j}x_3).
\fe

\section{Free field representation of $\mathsf{Vir}\oplus \mathsf{Vir}$ primaries}

\label{app:freefield}

The $\mathsf{Vir}\oplus \mathsf{Vir}$ primaries $v_n, v_n^\alpha$ can be directly constructed using a free field realization of the $\mathsf{SCA}_\nu \oplus \mathsf{F}_{\nu}$ algebra \cite{Hadasz:2013dza}. Using free boson operators $\left\{c_n | n\neq 0\right\}$, free fermion operators $\left\{\eta_r| r\in \mathbb{Z}+\nu\right\}$ and the central element $\hat{P}$ that satisfy the algebraic relations
\ie
[c_n,c_m] =n\delta_{m+n,0}, ~\left\{\eta_r,\eta_s\right\}= \delta_{r+s,0},
\fe
we can realize the generators of the superconformal algebra $\mathsf{SCA}_\nu$ as
\ie
L_{n\neq 0} & =\frac{1}{2} \sum_{k \neq 0, n} c_k c_{n-k}+\frac{1}{2} \sum_r r \eta_{n-r} \eta_r+\frac{i}{2}(Q n - 2 \hat{P}) c_n, \\
L_0 & =\sum_{k>0} c_{-k} c_k+\sum_{r>0} r \eta_{-r} \eta_r+\frac{1}{2}\left(\frac{Q^2}{4}-\hat{P}^2\right)+\frac{1-2\nu}{16}, \\
G_r & =\sum_{n \neq 0} c_n \eta_{r-n}+i(Q r - \hat{P}) \eta_r.
\fe
Each free boson mode $c_n$ and free fermion mode $\eta_r$ can be expressed directly in terms of $G_{-r}, L_{-n}$ and $\hat{P}$. 

It can be proved that $\mathcal{M}_{\NS}(h)$ is isomorphic to the Fock space generated by acting with $c_{n<0}, \psi_{r<0}, \eta_{r<0}$ on a ground state $\phi(P)$ that satisfies $\hat{P}\phi(P)=P\phi(P)$ \footnote{We take the branch of $P$ as follows: For $h$ smaller than $Q^2/8$, $P$ is taken to be the positive root of $\frac{1}{4}Q^2-2h$. For $h$ greater than $Q^2/8$, we take $P=ip$ such that $p>0$. In the R-sector, $P$ is taken to be the positive root of $2\beta^2$ for $\beta^2 >0$ and the root with positive imaginary part for $\beta^2 <0$}, and the $\mathsf{Vir}\oplus\mathsf{Vir}$ primaries in $\mathcal{M}_{\NS}(h)$ can be constructed using the free field operators $\chi_{r}(P) = \psi_r - i\eta_r(P)$ as
\ie \label{vNS}
v_n(P) = \chi_{-\frac{4n-1}{2}}(P)\chi_{-\frac{4n-3}{2}}(P)\cdots \chi_{-\frac{1}{2}}(P)\phi(P)
\fe
where $\eta_r(P)$ is the solution of $\eta_r$ in terms of $L_n, G_r, \hat{P}$ and with $\hat{P}$ replaced by $P$. $v_n$ with negative $n$ is defined by the reflection formula
\ie
v_n(P) = v_{-n}(-P).
\fe
Similarly in the R-sector, the $\mathsf{Vir}\oplus\mathsf{Vir}$ primaries in $\mathcal{M}_{\R}(\beta)$ can be constructed by
\ie \label{vR}
v_n^{2n+\frac{1}{2}\,\text{mod}\, 2}(P) &= \chi_0(P)\chi_{-1}(P)\cdots \chi_{-(2n-\frac{1}{2})}(P)(u^0\otimes w_\beta^+),\\
v_n^{2n-\frac{1}{2}\,\text{mod}\,2}(P) &= \chi_0(P)\chi_{-1}(P)\cdots \chi_{-(2n-\frac{1}{2})}(P)\chi_0(-P)(u^0\otimes w_\beta^+),
\fe
for $n>0$, and by the reflection formula
\ie
(v^\alpha_{n}(P))_+ = (v_{-n}^\alpha(-P))_+,\quad (v^\alpha_{n}(P))_- = -(v_{-n}^\alpha(-P))_-,
\fe
for $n<0$. The subscripts $\pm$ denote the components ending in \(w_\beta^+\) and \(w_\beta^-\), respectively. 

We here list some low-level $\mathsf{Vir\oplus Vir}$ primaries:
\ie \label{expVVprim}
v_0(P)&= \mathbf{1}\otimes \phi(P)\\
v_{\pm \frac{1}{2}}(P) &= \psi_{-\frac{1}{2}}\mathbf{1}\otimes \phi(P) + \frac{1}{Q/2\pm P}\mathbf{1}\otimes G_{-\frac{1}{2}}\phi(P),\\
v^0_{\pm \frac{1}{4}}(P) &= u^0\otimes w_\beta^+\mp e^{i\pi/4} u^1\otimes w^-_\beta,\\
v^1_{\pm \frac{1}{4}}(P) &= \frac{1}{\sqrt{2}}u^1\otimes w_\beta^+\pm \frac{1}{\sqrt{2}} e^{i\pi/4} u^0\otimes w^-_\beta.
\fe

The BPZ norms of these $\mathsf{Vir}\oplus \mathsf{Vir}$ primaries are computed in \cite{Hadasz:2013dza}. We define the function $\ell(x,m)$ to be
\ie
 \ell(x,m)=2^{(1-(-1)^m)/16}
 \prod_{\substack{r,s\ge0,\ r+s<m\\r+s\equiv m\ (\mathrm{mod}\,2)}}
 (x+rb+s/b),\qquad m\ge0,
\fe
and $\ell(x,0)=1$. For $m<0$, we define their $\ell(x,m)$ using the reflection formula
\ie
 \ell(x,-m)=
 \begin{cases}
 \ell(Q-x,m),&m>0\text{ odd},\\
 (-1)^{m/2}\ell(Q-x,m),&m>0\text{ even}.
 \end{cases}
\fe
The BPZ norms of the $\mathsf{Vir}\oplus \mathsf{Vir}$ primaries with $n>0$ are given by
\ie
\|v_n(P)\|^2 &= i^{2n\text{ mod }2}(-1)^{2n(1+\epsilon(\phi))} 2^{2n} \frac{\ell(2P,4n)}{\ell(Q+2P, 4n)},\\
\|v_n^0(P)\|^2&=2^{2\lfloor M/2\rfloor+1}
               \frac{\ell(2P,4n)}{\ell(Q+2P,4n)},\\
 \|v_n^1(P)\|^2&=(-i)2^{2\lceil M/2\rceil}
               \frac{\ell(2P,4n)}{\ell(Q+2P,4n)},
\fe
where $M = 2n-\frac{1}{2}$, $\epsilon(\phi)$ is the Grassmann parity of the superconformal primary $\phi(P)$. For the norms of the $\mathsf{Vir}\oplus \mathsf{Vir}$ primaries with $n<0$, one replaces $P$ with $-P$, $\beta$ with $-\beta$, $n$ with $-n$ and uses the formula above.

The (NS, NS, NS) branching coefficient is also fixed in \cite{Hadasz:2013dza} using the free field representation. In our convention, their result reads
\ie \label{NSbranching}
\mathbf{B}_a&(n_1, n_2, n_3) = \frac{(-1)^{2n_3(1+\epsilon(\phi_2))}(-i)^ae^{\frac{i\pi}{4}(\epsilon_1-\epsilon_2-\epsilon_3)}}{\prod_{i=1}^3\sqrt{\ell(2P_i, 4n_i)\ell(Q+2P_i, 4n_i)}}\\
&\quad \times \ell\left(\tfrac{Q}{2}+P_1+P_2+P_3, 2(n_1+n_2+n_3)\right)\ell\left(\tfrac{Q}{2}-P_1+P_2+P_3, 2(-n_1+n_2+n_3)\right)\\
&\quad \times \ell\left(\tfrac{Q}{2}+P_1+P_2-P_3, 2(n_1+n_2-n_3)\right) \ell\left(\tfrac{Q}{2}-P_1+P_2-P_3, 2(-n_1+n_2-n_3)\right),
\fe
where $\epsilon_i = \epsilon(v_{i,n_i})-\epsilon(\phi_i)$ is treated as the difference between the two integers $\epsilon(\phi_i), \epsilon(v_{i,n_i})\in \{0,1\}$. This formula only applies when $\epsilon_1+\epsilon_2+\epsilon_3\equiv a \mod 2$. In the code, we used this formula directly for the (NS, NS, NS) branching coefficient, while the (NS, R, R) ones are obtained from the $L_1$ Ward identities. 

The first few (NS, R, R) branching coefficients can be computed directly from \eqref{expVVprim}. For $n_2, n_3 = \pm \frac{1}{4}, \alpha_2, \alpha_3=0,1$, one has
\ie
\mathbb B_0^{(\sigma)}(0,n_2,n_3,\alpha_2,\alpha_3)
&=\frac{1+16\sigma n_2n_3}{2}
(-1)^{\alpha_2}\delta_{\alpha_2,\alpha_3},\\
\mathbb B_1^{(\sigma)}(0,n_2,n_3,\alpha_2,\alpha_3)
&=2(n_3-\sigma n_2)\,
i^{\,1-\alpha_2}\delta_{\alpha_2,\,1-\alpha_3},\\
\mathbb B_f^{(\sigma)}
 \left(\pm\tfrac12,n_2,n_3,\alpha_2,\alpha_3\right)
&=-2n_3(-i)^{\,f+\epsilon(\phi_1)}
\frac{\frac Q2\pm P_1+4n_2P_2+4n_3P_3}
     {\sqrt{\pm P_1\left(\frac Q2\pm P_1\right)}}
\,\mathbb B_{1-f}^{(\sigma)}
 (0,n_2,n_3,\alpha_2,\alpha_3).
\fe
As we have seen in section \ref{sec:blockdecomp}, these low-level branching coefficients, together with the recursion relation provided by the $L_1$ Ward identities, determines all (NS, R, R) branching coefficients.

We now prove that, for $n>0$, $L_1v_n$ and $L_1v_n^\alpha$ contain only level-$(4n-3)$ descendants of $v_{n-1}$ and $v_{n-1}^\alpha$, respectively. Likewise, $L_{-1}v_n$ and $L_{-1}v_n^\alpha$ contain only level-1 descendants in their original $\mathsf{Vir}\oplus\mathsf{Vir}$ modules and level-$(4n-1)$ descendants in the modules generated by $v_{n-1}$ and $v_{n-1}^\alpha$, respectively.

We prove the statement in the R sector, assuming $n>0$ throughout. The NS case is analogous. Let $\mathcal{V}^{\alpha}_n:=V_{c^{(1)}}(h_n^{(1)})\otimes V_{c^{(2)}}(h_n^{(2)})$ denote the copy of the double Virasoro module generated by $v_n^\alpha$. We retain the assumptions $b\neq0,\pm1$ and $P\notin\frac12(b\mathbb Z+b^{-1}\mathbb Z)$ used in the branching decomposition.

Our starting point is the following operator equation
\ie
(1-b^2)\bigl((L_{-1}^{(1)})^2\psi\bigr)(z)
+2b^2(L_{-2}^{(1)}\psi)(z)=0,
\fe
which follows from the vanishing of the corresponding descendant of $\ket{\psi}=\psi_{-1/2}\ket{\Omega}$, where $\ket{\Omega}$ is the NS identity vacuum. Here, the modes acting on $\psi$ define local descendants at the insertion point; the resulting field identity also holds on R modules. Taking its matrix element between double Virasoro primaries gives
\ie
0&=(1-b^2)\langle\langle v_m^{1-\alpha}|
\bigl((L_{-1}^{(1)})^2\psi\bigr)(1)|v_n^\alpha\rangle
+2b^2
\langle\langle v_m^{1-\alpha}|
(L_{-2}^{(1)}\psi)(1)
|v_n^\alpha\rangle\\
{}&
=(1-b^2)
\bigl(h_m^{(1)}-h_{n+\frac12}^{(1)}\bigr)
\bigl(h_m^{(1)}-h_{n-\frac12}^{(1)}\bigr) \langle\langle v^{1-\alpha}_m|\psi(1)|v^{\alpha}_n\rangle.
\fe
Since $b\neq\pm1$, a nonzero overlap $\langle\langle v_m^{1-\alpha}|\psi(1)|v_n^\alpha\rangle$ requires $h_m^{(1)}=h_{n\pm\frac12}^{(1)}$. The genericity assumption on $P$ excludes the reflected solutions, so $m=n\pm\frac12$. The Virasoro Ward identities determine all descendant matrix elements from these primary overlaps. Together with the nondegeneracy of the BPZ pairings, this gives the selection rule
\ie\label{eq:selectionrule}
\psi_r\! \mathcal{V}^{\alpha}_n
\subseteq
 \mathcal{V}^{1-\alpha}_{n+\frac{1}{2}}
\oplus
 \mathcal{V}^{1-\alpha}_{n-\frac{1}{2}}.
\fe
The Virasoro generator $L_k$ can be represented as 
\ie
L_k=L_k^{(1)}+L_k^{(2)}-L_k^{\mathsf F} \quad k\in \mathbb{Z}.
\fe
The generators $L_k^{(1)}$ and $L_k^{(2)}$ preserve each double Virasoro module, while $L_k^{\mathsf F}$ in \eqref{eq:fermionVirasoro} is quadratic in the fermion modes, apart from its scalar term at $k=0$. Applying \eqref{eq:selectionrule} twice therefore gives
\ie
L_k \mathcal{V}_n^\alpha \subseteq
 \mathcal{V}^{\alpha}_{n+1}
\oplus
 \mathcal{V}^{\alpha}_{n-1}
 \oplus
 \mathcal{V}^{\alpha}_{n}.
\fe
To complete the proof, we combine this selection rule with level counting. Since
\ie
[L_0^{(1)}+L_0^{(2)},L_k]=[L_0+L_0^{\mathrm F},L_k]=-kL_k,
\fe
the state $L_kv_n^\alpha$ lies in the eigenspace of $L_0^{(1)}+L_0^{(2)}$ with eigenvalue $h_n^{(1)}+h_n^{(2)}-k$. Consequently, its component in $\mathcal V_m^\alpha$ must have descendant level
\ie
h_n^{(1)}+h_n^{(2)}-k-h_m^{(1)}-h_m^{(2)}=2n^2-2m^2-k.
\fe

For $k=1$, the three allowed branches $m=n-1,n,n+1$ therefore correspond to levels $4n-3$, $-1$, and $-4n-3$, respectively. Since descendant levels are nonnegative and $n>0$, only the $m=n-1$ branch can contribute. Thus, $L_1v_n^\alpha$ consists entirely of level-$(4n-3)$ descendants of $v_{n-1}^\alpha$; it vanishes if $4n-3<0$.

Similarly, for $k=-1$, the three branches correspond to levels $4n-1$, $1$, and $-4n-1$, respectively. The $m=n+1$ branch is again excluded. Hence, $L_{-1}v_n^\alpha$ contains only level-1 descendants of $v_n^\alpha$ and level-$(4n-1)$ descendants of $v_{n-1}^\alpha$.

This completes the proof in the R sector. The NS sector follows by the same argument, as does the analogous statement for $n<0$.

\section{The cross-channel test}

\label{app:Zratiocheck}

In this appendix, we present the details of the cross-channel check mentioned in Section \ref{sec:numerics}. 

\paragraph{The super-Liouville partition function} We consider the $\mathcal{N}=1$ super-Liouville theory with $Q=b+b^{-1}$, whose central charge is $c=\frac{3}{2}(1+2Q^2)$. The spectrum of this theory is given by
\ie
h_\NS(p) = \frac{Q^2}{8}+\frac{p^2}{2},\quad h_{\R}(p)= \frac{Q^2}{8}+\frac{p^2}{2} + \frac{1}{16},
\fe
with $p$ ranging from $0$ to $\infty$. This corresponds to $P=ip$ and $\beta = ip/\sqrt{2}$ in the convention used above.

The (NS, NS, NS) structure constants $C^{(a)}(p_1, p_2, p_3)$ are given by \cite{Rashkov:1996np, Poghossian:1996agj}
\ie
C^{(0)}(p_1,p_2,p_3)&=\frac{\Upsilon'_{\NS}(0)}{b^3}
 \frac{\prod_{j=1}^3
       \sqrt{\Upsilon_{\NS}(2ip_j)\Upsilon_{\NS}(-2ip_j)}}
 {\Upsilon_{\NS}(Q/2+i\sum_jp_j)
       \prod_{j=1}^3\Upsilon_{\NS}(Q/2+i\sum_kp_k-2ip_j)},\\
C^{(1)}(p_1,p_2,p_3)&=\frac{2i\Upsilon'_{\NS}(0)}{b^3}
 \frac{\prod_{j=1}^3
       \sqrt{\Upsilon_{\NS}(2ip_j)\Upsilon_{\NS}(-2ip_j)}}
 {\Upsilon_{\R}(Q/2+i\sum_jp_j)
       \prod_{j=1}^3\Upsilon_{\R}(Q/2+i\sum_kp_k-2ip_j)},
\fe
where the Upsilon function $\Upsilon$ is defined for $0<\mathrm{Re}(x)<Q$ by
\ie
\log\Upsilon_b(x)=\int_0^\infty\frac{dt}{t}
 \left[\left(\frac Q2-x\right)^2e^{-t}
 -\frac{\sinh^2((Q/2-x)t/2)}
 {\sinh(bt/2)\sinh(t/(2b))}\right],
\fe
and the functions $\Upsilon_\NS, \Upsilon_\R$ are defined as
\ie
\Upsilon_{\NS}(x)=\Upsilon_b(x/2)\Upsilon_b((x+Q)/2),\quad
 \Upsilon_{\R}(x)=\Upsilon_b((x+b)/2)\Upsilon_b((x+b^{-1})/2).
\fe
Specifically, one has $\Upsilon'_{\NS}(0)=\frac12 \Upsilon_b(b)$.

The (NS, R, R) structure constants $C^{(f,\sigma)}$ are given by
\ie
C^{(0,+)}(p_1,p_2,p_3)
 &=\frac{\Upsilon'_{\NS}(0)}{2b^3}
 \frac{\sqrt{\prod_{\sigma=\pm1}
     \Upsilon_{\NS}(2i\sigma p_1)
     \Upsilon_{\R}(2i\sigma p_2)\Upsilon_{\R}(2i\sigma p_3)}}
 {\prod_{\sigma=\pm1}
     \Upsilon_{\R}(Q/2+i(p_2+p_3+\sigma p_1))
     \Upsilon_{\NS}(Q/2+i(p_1+\sigma(p_2-p_3)))},\\
 C^{(0,-)}(p_1,p_2,p_3)
&=\frac{\Upsilon'_{\NS}(0)}{2b^3}
 \frac{\sqrt{\prod_{\sigma=\pm1}
     \Upsilon_{\NS}(2i\sigma p_1)
     \Upsilon_{\R}(2i\sigma p_2)\Upsilon_{\R}(2i\sigma p_3)}}
 {\prod_{\sigma=\pm1}
     \Upsilon_{\NS}(Q/2+i(p_2+p_3+\sigma p_1))
     \Upsilon_{\R}(Q/2+i(p_1+\sigma(p_2-p_3)))}.
\fe
The other two coefficients $C^{(1,\sigma)}$ are given by $C^{(1,\sigma)}=iC^{(0,-\sigma)}$.

We will focus on the genus two partition function of this theory in the genus two theta channel. In the theta channel with NS-NS-NS edges, the $a$ label in both 3-point functions $\rho_a$ is forced to be the same by the Gram matrices. We can then denote the all-NS theta channel superconformal block as $\mathbf{F}^{(a)}(h_{1,2,3};q_{1,2,3},\eta_{1,2,3})$. For the theta channel with NS-R-R edges, the two $f$ labels of $\rho^{(\sigma)}_f$ are forced to be the same, but the two $\sigma$ labels can be different. We will denote the NS-R-R theta channel block as $\mathbf{F}^{(f,\sigma,\sigma')}(h_1, \beta_{2,3};q_{1,2,3},\eta_{1,2,3})$, where $\sigma, \sigma'$ are the $\sigma$-labels of the 3-point functions on the $z_1$ and the $z_2$ sphere. There is an odd Ramond loop in this channel if $\sigma \sigma'=-1$. Using these blocks, the super-Liouville partition function can be written as
\ie
Z_{\mathrm{SL}}(q_{1,2,3}, \eta_{1,2,3}) = \sum_{a=0}^1 (-1)^a \int_0^\infty \prod_{e=1}^3 \frac{dp_e}{\pi} |q_e^{h_e}|^2 C^{(a
)}(p_1,p_2,p_3)^2 |\mathbf{F}^{(a)}(h_{1,2,3};q_{1,2,3},\eta_{1,2,3})|^2
\fe
in the NS-NS-NS theta channel, where the factor $(-1)^a$ is given by the sign $(-1)^{S_\Gamma}$ as defined in \eqref{sgamma}. In the NS-R-R theta channel, one has
\ie
Z_{\mathrm{SL}}(q_{1,2,3}, \eta_{1,2,3}) = \sum_{f=0}^1(-1)^f\sum_{\sigma,\sigma'=\pm} \int_0^\infty \prod_{e=1}^3 \frac{dp_e}{\pi} |q_e^{h_e}|^2 &C^{(f,\sigma)}(p_1,p_2,p_3)C^{(f,\sigma')}(p_1,p_2,p_3)\\
&\times |\mathbf{F}^{(f,\sigma, \sigma')}(h_{1},\beta_{2,3};q_{1,2,3},\eta_{1,2,3})|^2.
\fe

\paragraph{The free theory partition function} The free boson + fermion theory partition function can be directly computed in terms of the plumbing parameters using the formula \cite{Yin:2007gv}
\ie
Z_{\mathrm{free}}(q_e, \eta_e) = \frac{|P(q_e)|^3 \left|\theta[\alpha(\eta_e)| \beta(\eta_e)](\Omega(q_e))\right|}{\sqrt{\det 2 \operatorname{Im}\Omega(q_e)}},
\fe
where $\Omega(q_e)$ is the period matrix of the sewn Riemann surface. The theta function $\theta[\alpha|\beta]$ is defined as
\ie
\theta[\alpha|\beta](\Omega) = \sum_{n\in \mathbb{Z}^2}\exp\left[i\pi(n+\tfrac{\alpha}{2})^\mathrm{T}\Omega(n+\tfrac{\alpha}{2})+i\pi(n+\tfrac{\alpha}{2})^\mathrm{T}\beta\right].
\fe
In the genus two theta channel, the characteristics $\alpha, \beta$ are related to the spin lift signs $\eta_e$ by
\ie
&\text{NS-NS-NS:}\quad &\alpha(\eta_e) = (0,0),\quad \beta(\eta_e)=\left(\tfrac{1}{2}(1+\eta_1\eta_3),\tfrac{1}{2}(1+\eta_1\eta_2)\right),\\
&\text{NS-R-R:}\quad &\alpha(\eta_e) = (1,1),\quad \beta(\eta_e)=\left(\tfrac{1}{2}(1-\eta_1\eta_3),\tfrac{1}{2}(1+\eta_1\eta_2)\right).
\fe

Finally, the function $P(q_e)$ is given by the following. The Schottky group of the sewn Riemann surface is a free group generated by the Mobius transformations
\ie
g_0(z) = q_1q_3z,\quad g_1(z) = \frac{q_1(q_2-1)z+1}{1-q_1z}.
\fe
For each primitive word $\gamma$ in the Schottky group, let $M_\gamma$ be the corresponding $SL(2,\mathbb{C})$ transformation, and we define
\ie
k_\gamma(q_e) = \frac{\operatorname{tr}M_\gamma-
  \sqrt{(\operatorname{tr}M_\gamma)^2-4\det M_\gamma}}
 {\operatorname{tr}M_\gamma+
  \sqrt{(\operatorname{tr}M_\gamma)^2-4\det M_\gamma}},
\fe
where the branch of the square roots is specified by requiring $|k_\gamma|<1$. The function $P(q_e)$ is then defined by
\ie
P(q_e) = \prod_{[\gamma]} \prod_{m=1}^\infty  \frac{1}{\left(1-k_\gamma(q_e)^m\right)^{\frac{1}{2}}},
\fe
where the product goes over all conjugacy classes $[\gamma]$ of primitive words in the Schottky group.

\paragraph{The cross-channel test} The pairs of plumbing parameters $\{q_e^{\NS\NS\NS}, \eta_e^{\NS\NS\NS}\}, \{q_e^{\NS\R\R}, \eta_e^{\NS\R\R}\}$ in the NS-NS-NS and the NS-R-R theta channels used in the test are listed in table \ref{qtable}. One can check that they give equivalent period matrices up to a symplectic change of basis using the algorithm in \cite{Christian:new}. 

For each Riemann surface specified by $\{q_e\}$, we compute the following quantity $\mathcal{Q}$ using the algorithms described above for two different spin structures given by the two shown sets of $\eta_e$:
\ie
\mathcal{Q}=\frac{Z_\mathrm{\mathrm{SL}}(q_e^{\NS\R\R}, \eta_e^{\NS\R\R})}{Z_\mathrm{\mathrm{SL}}(q_e^{\NS\NS\NS}, \eta_e^{\NS\NS\NS})} \left(\frac{Z_{\mathrm{free}}(q_e^{\NS\NS\NS}, \eta_e^{\NS\NS\NS})}{Z_{\mathrm{free}}(q_e^{\NS\R\R}, \eta_e^{\NS\R\R})}\right)^{1+2Q^2}
\fe
The super-Liouville theory is taken to be that of $b=\frac{7}{5}$. The superconformal blocks are computed to the coefficients of $q_1^{n_1}q_2^{n_2}q_3^{n_3}$ up to total level $n_1+n_2+n_3=8$ for the NS-NS-NS theta channel and total level $5$ in the NS-R-R theta channel. Across the 20 spin Riemann surfaces tested, the maximum deviation of $\mathcal{Q}$ from 1 is $4.03\times10^{-5}$. 

{\scriptsize
\renewcommand{\arraystretch}{1.10}
\setlength{\tabcolsep}{2.2pt}
\begin{longtable}{@{}c|r|r|r|c c|c c@{}}
\hline\hline
Channel & $q_1$ & $q_2$ & $q_3$ & \multicolumn{2}{c|}{First spin structure} & \multicolumn{2}{c}{Second spin structure}\\
 & & & & $(\eta_1, \eta_2, \eta_3)$ & $|\mathcal{Q}-1|$ & $(\eta_1, \eta_2, \eta_3)$ & $|\mathcal{Q}-1|$\\
\hline\hline
\endfirsthead
\hline\hline
Channel & $q_1$ & $q_2$ & $q_3$ & \multicolumn{2}{c|}{First spin structure} & \multicolumn{2}{c}{Second spin structure}\\
 & & & & $(\eta_1, \eta_2, \eta_3)$ & $|\mathcal{Q}-1|$ & $(\eta_1, \eta_2, \eta_3)$ & $|\mathcal{Q}-1|$\\
\hline\hline
\endhead
$\NS\NS\NS$ & $-0.071250-0.042549i$ & $-0.019696+0.005144i$ & $+0.017998-0.080979i$ & $(-,-,-)$ & $1.57\times10^{-5}$ & $(-,-,+)$ & $2.56\times10^{-5}$\\*
$\NS\R\R$ & $-0.027941+0.071218i$ & $-0.057774+0.066640i$ & $-0.026376-0.017260i$ & $(+,-,+)$ &  & $(-,-,+)$ & \\[2pt]
\hline
$\NS\NS\NS$ & $-0.082935+0.006449i$ & $-0.033283-0.013330i$ & $+0.049890-0.014666i$ & $(-,-,+)$ & $1.58\times10^{-5}$ & $(-,-,-)$ & $5.30\times10^{-6}$\\*
$\NS\R\R$ & $-0.060523+0.011892i$ & $-0.027736+0.021163i$ & $-0.043145-0.012892i$ & $(+,-,+)$ &  & $(-,-,+)$ & \\[2pt]
\hline
$\NS\NS\NS$ & $-0.062544+0.013403i$ & $-0.015537-0.007620i$ & $+0.020629-0.007765i$ & $(-,-,+)$ & $2.59\times10^{-5}$ & $(-,-,-)$ & $2.98\times10^{-5}$\\*
$\NS\R\R$ & $-0.016109+0.005454i$ & $-0.039397+0.032408i$ & $-0.072252-0.037339i$ & $(+,-,+)$ &  & $(-,-,+)$ & \\[2pt]
\hline
$\NS\NS\NS$ & $-0.080397+0.073055i$ & $-0.032588-0.026727i$ & $+0.023875-0.010459i$ & $(-,-,+)$ & $3.99\times10^{-6}$ & $(-,-,-)$ & $4.13\times10^{-6}$\\*
$\NS\R\R$ & $-0.029248+0.019339i$ & $-0.011178+0.008110i$ & $-0.056249-0.036215i$ & $(+,-,+)$ &  & $(-,-,+)$ & \\[2pt]
\hline
$\NS\NS\NS$ & $-0.057225-0.017265i$ & $-0.017256+0.000446i$ & $+0.073492-0.059010i$ & $(-,-,-)$ & $4.03\times10^{-5}$ & $(-,-,+)$ & $2.22\times10^{-5}$\\*
$\NS\R\R$ & $-0.061589+0.039921i$ & $-0.069915+0.051999i$ & $-0.046420-0.019395i$ & $(+,-,+)$ &  & $(-,-,+)$ & \\[2pt]
\hline
$\NS\NS\NS$ & $-0.069370+0.024053i$ & $-0.020448-0.018347i$ & $+0.013474-0.017942i$ & $(-,-,+)$ & $2.08\times10^{-6}$ & $(-,-,-)$ & $1.84\times10^{-6}$\\*
$\NS\R\R$ & $-0.013911+0.016093i$ & $-0.023198+0.022171i$ & $-0.071563-0.025133i$ & $(+,-,+)$ &  & $(-,-,+)$ & \\[2pt]
\hline
$\NS\NS\NS$ & $-0.074203+0.022952i$ & $-0.037505-0.020450i$ & $+0.040852-0.011942i$ & $(-,-,+)$ & $1.31\times10^{-5}$ & $(-,-,-)$ & $8.93\times10^{-6}$\\*
$\NS\R\R$ & $-0.049364+0.012976i$ & $-0.022885+0.012734i$ & $-0.050842-0.011843i$ & $(+,-,+)$ &  & $(-,-,+)$ & \\[2pt]
\hline
$\NS\NS\NS$ & $-0.067061+0.003058i$ & $-0.033476-0.018628i$ & $+0.040470-0.032913i$ & $(-,-,+)$ & $1.89\times10^{-5}$ & $(-,-,-)$ & $1.00\times10^{-5}$\\*
$\NS\R\R$ & $-0.048123+0.028738i$ & $-0.030295+0.020925i$ & $-0.050247-0.004106i$ & $(+,-,+)$ &  & $(-,-,+)$ & \\[2pt]
\hline
$\NS\NS\NS$ & $-0.078821+0.015521i$ & $-0.025878-0.005284i$ & $+0.041879-0.035882i$ & $(-,-,+)$ & $7.22\times10^{-6}$ & $(-,-,-)$ & $6.30\times10^{-6}$\\*
$\NS\R\R$ & $-0.040796+0.039739i$ & $-0.034555+0.023510i$ & $-0.044637-0.028816i$ & $(+,-,+)$ &  & $(-,-,+)$ & \\[2pt]
\hline
$\NS\NS\NS$ & $-0.089186+0.052432i$ & $-0.033207-0.014493i$ & $+0.016411-0.010169i$ & $(-,-,+)$ & $3.92\times10^{-6}$ & $(-,-,-)$ & $4.13\times10^{-6}$\\*
$\NS\R\R$ & $-0.018493+0.016088i$ & $-0.017174+0.011483i$ & $-0.045417-0.035425i$ & $(+,-,+)$ &  & $(-,-,+)$ & \\
\hline\hline
\caption{Plumbing parameters and spin-lift signs used in the cross-channel test. Each row of this table correspond to a genus-2 Riemann surface, and the plumbing parameters $q_1,q_2,q_3$ used to construct this Riemann surface in the NS-NS-NS and the NS-R-R theta channel are listed in the first three colomns. For each Riemann surface, we compute the quantity $|\mathcal{Q}-1|$ for two different spin structures on the surface. Their spin lift signs $\eta_1,\eta_2,\eta_3$ and the resulting quantities $|\mathcal{Q}-1|$ are listed in the last two colomns.}
\label{qtable}
\end{longtable}
}

\end{appendix}

\printbibliography
\end{document}